\documentclass[12pt]{iopart}

\usepackage{graphicx}
\usepackage{siunitx}
\usepackage{units}
\usepackage{overpic}
\usepackage{float}
\usepackage{hyperref}
\usepackage{framed, color}
\usepackage{bbm}
\usepackage[dvipsnames]{xcolor}

\usepackage{amssymb}

\usepackage{tabularx}

\newcommand*\diff{\mathop{}\!\mathrm{d}}

\usepackage[numbers,square,comma,sort&compress]{natbib}

\usepackage{enumitem} % To customize lists better

\usepackage{color}
\usepackage{url}

\usepackage{mleftright} %to get rid of the extra space around parentheses with \left and \right: write \mleft, \mright instead 

\usepackage{hyperref}

\newcommand{\figref}[1]{\mbox{Fig.~\ref{#1}}}
\newcommand{\tabref}[1]{\mbox{Table~\ref{#1}}}
\newcommand{\secref}[1]{\mbox{Sec.~\ref{#1}}}

\newcommand{\eqref}[1]{\mbox{Eq.~(\ref{#1})}}

\newcommand{\figpanel}[2]{Fig.~\hyperref[#1]{\ref*{#1}(#2)}}
\newcommand{\figpanelNoPrefix}[2]{\hyperref[#1]{\ref*{#1}(#2)}}

\newcommand{\bra}[1]{\langle #1|}
\newcommand{\ket}[1]{|#1\rangle}
\newcommand{\braket}[2]{\langle #1|#2\rangle}
\newcommand{\ketbra}[2]{\mleft| #1 \rangle \langle #2 \mright|}
\newcommand{\brakket}[3]{\mleft\langle #1\mleft| #2 \mright| #3\mright\rangle}
\newcommand{\expec}[1]{\mleft\langle #1 \mright\rangle}

\newcommand{\comm}[2]{\mleft[ #1, #2 \mright]}

\newcommand{\abs}[1]{\mleft|#1\mright|}

\newcommand{\abssq}[1]{\mleft| #1 \mright|^2}

\newcommand{\nn}{\nonumber}

\newcommand{\be}{\begin{equation}}
\newcommand{\ee}{\end{equation}}
\newcommand{\bea}{\begin{eqnarray}}
\newcommand{\eea}{\end{eqnarray}}

\AtBeginDocument{%
    \newwrite\bibnotes
    \def\bibnotesext{Notes.bib}
    \immediate\openout\bibnotes=\jobname\bibnotesext
    \immediate\write\bibnotes{@CONTROL{REVTEX41Control}}
    \immediate\write\bibnotes{@CONTROL{%
    apsrev41Control,author="08",editor="1",pages="0",title="0",year="1"}}
     \if@filesw
     \immediate\write\@auxout{\string\citation{apsrev41Control}}%
    \fi
}%

\begin{document}

\title{Ultimate quantum limit for amplification: a single atom in front of a mirror}

\author{Emely Wiegand$^1$, Ping-Yi Wen$^2$, Per Delsing$^1$, Io-Chun Hoi$^{2,3}$, Anton Frisk Kockum$^1$}

\address{$^1$ Department of Microtechnology and Nanoscience (MC2), Chalmers University of Technology, 412 96 Gothenburg, Sweden}
\address{$^2$ Department of Physics, National Tsing Hua University, Hsinchu 30013, Taiwan}
\address{$^3$ Center for Quantum Technology, National Tsing Hua University, Hsinchu 30013, Taiwan}
\ead{wiegand@chalmers.se}
\ead{anton.frisk.kockum@chalmers.se}

\begin{abstract}

We investigate three types of amplification processes for light fields coupling to an atom near the end of a one-dimensional semi-infinite waveguide. We consider two setups where a drive creates population inversion in the bare or dressed basis of a three-level atom and one setup where the amplification is due to higher-order processes in a driven two-level atom. In all cases, the end of the waveguide acts as a mirror for the light. We find that this enhances the amplification in two ways compared to the same setups in an open waveguide. Firstly, the mirror forces all output from the atom to travel in one direction instead of being split up into two output channels. Secondly, interference due to the mirror enables tuning of the ratio of relaxation rates for different transitions in the atom to increase population inversion. We quantify the enhancement in amplification due to these factors and show that it can be demonstrated for standard parameters in experiments with superconducting quantum circuits.

\end{abstract}
%\keywords{Amplification, Circuit QED, Quantum Optics}
\maketitle

%%%%%%%%%%%%%%%%%%%%%%%%%%%%%%%%%%%%%%%%%%%%%%%

\section{Introduction}

Amplification of measurement signals is crucial to achieve good signal-to-noise ratios in many experiments in quantum information and quantum optics~\cite{Clerk2010, Aumentado2020}. Ideally, amplifiers used for such tasks should be compact, add as little noise as possible~\cite{Haus1962, Caves1982}, and produce high gain. To reach the ultimate limit in terms of size, a single atom or other quantum emitter could be used as an amplifier. However, to achieve high gain with a single quantum emitter coupled to an electromagnetic field in free space is extremely challenging, since imperfect spatial mode matching leads to a weak coupling~\cite{Leuchs2013, Gerhardt2007, Vamivakas2007, Wrigge2008, Tey2008, Hwang2009, Leong2016}. The mode matching, and thus a strong coupling, is much easier to achieve when the propagation of the field is confined to a one-dimensional (1D) waveguide. Such systems are widely studied in waveguide quantum electrodynamics (waveguide QED), which has proven an excellent platform for quantum-optical experiments~\cite{Roy2017, Gu2017}.

In the past two decades, many quantum-optics phenomena have been demonstrated using superconducting circuits~\cite{You2011, Gu2017, Kockum2019a, Blais2020}, e.g., lasing~\cite{Astafiev2007, Ashab2009, You2007, Marthaler2011}. Superconducting circuits consist of superconducting qubits~\cite{Kockum2019a, Kjaergaard2020} coupled to a coplanar waveguide (either open or made into a resonator)~\cite{Gu2017, Blais2004, Wallraff2004, Astafiev2010_2} or three-dimensional cavities~\cite{Gu2017, Blais2020, Paik2011}. One advantage of superconducting circuits over natural atomic systems is that strong, and even ultrastrong, coupling between the quantum emitters and cavities or open waveguides can be achieved quite easily~\cite{Devoret2007, Bourassa2009, Niemczyk2010, Forn-Diaz2017, Yoshihara2017, Kockum2019, Forn-Diaz2019}. This advantage has, for example, been demonstrated by Wen et al.~\cite{Wen2018}, who used superconducting circuits to realize a \unit[7]{\%} amplification of a weak probe signal on a strongly-driven two-level system coupled to a waveguide. Similar experiments with many natural atoms~\cite{Wu1977} or a single quantum dot~\cite{Xu2007} were only able to achieve \unit[0.4]{\%} and \unit[0.005]{\%} amplification, respectively.

The vast majority of waveguide-QED experiments with superconducting circuits so far were performed with one or more superconducting qubits coupled to an \textit{open} waveguide~\cite{Gu2017, Roy2017, Astafiev2010_2, Astafiev2010, Hoi2011, Hoi2012, vanLoo2013, Koshino2013, Hoi2013, Hoi2013_2, Forn-Diaz2017, Liu2017, Mirhosseini2018, Sundaresan2019, Wen2019, Mirhosseini2019, Kannan2020, Vadiraj2020}. However, the waveguide can also be shorted or left open at one end, where an incoming electromagnetic field will be reflected with a phase shift~\cite{Johansson2009, Wilson2011, Hoi2015, Wen2018}. When a superconducting qubit is included~\cite{Hoi2015, Wen2018}, this setup is equivalent to putting an atom in front of a \textit{mirror}, which has been studied experimentally~\cite{Eschner2001, Wilson2003, Dubin2007, Hoi2015, Wen2018, Wen2019, Lu2020, Scigliuzzo2020} and theoretically~\cite{Meschede1990, Dorner2002, Beige2002, Dong2009, Koshino2012, Wang2012, Tufarelli2013, Fang2015, Shi2015, Pichler2016, Pichler2017, Wiegand2020, Wiegand2020a} for both natural and artificial atoms. In this article, we investigate the advantages of using an atom in front of a mirror, instead of an atom in an open waveguide, for signal amplification.

There are several ways to achieve amplification in an atomic system driven by an electromagnetic field. One amplification mechanism is population inversion, where excitations are pumped into higher atomic levels with a finite life time, where they stay long enough to induce amplification through stimulated emission~\cite{Sargent1974, Silfvast1996}. There are also mechanisms that can lead to amplification and lasing without inversion in the bare-state basis~\cite{Mompart2000}. For instance, if an atom is driven strongly, the energy levels can split and population inversion can occur in the dressed-state basis~\cite{Haroche1972, Koshino2013} if the drive is off resonance. If the drive is on resonance, the power spectrum exhibits the so-called Mollow triplet~\cite{Mollow1969, Astafiev2010_2, Abdumalikov2011, Lu2020}; amplification without population inversion can then be achieved at frequencies between the triplet peaks due to higher-order processes between the dressed states of the driven atom~\cite{Mollow1972, Wu1977, Friedmann1987, Xu2007, Wen2018}.

%===================================================
\begin{figure}[t]
	\centering
	\begin{minipage}{0.49\linewidth}
		\begin{overpic}[width=1\linewidth]{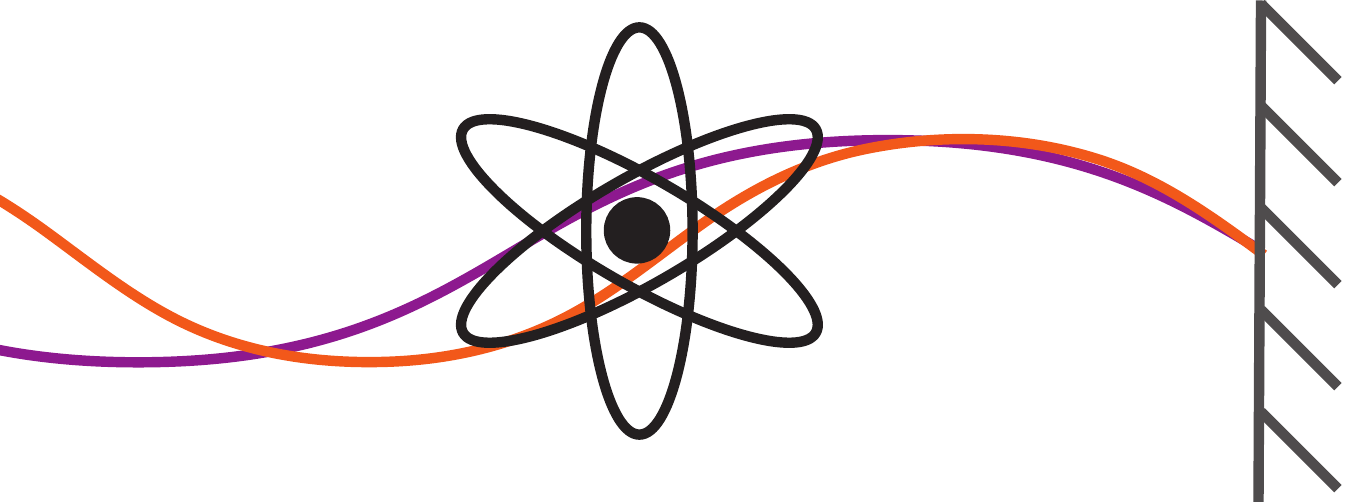}
			\put(-3,30){$\text{(a)}$}
			\put(101,30){$\text{(b)}$}
			\put(-3,-12){$\text{(c)}$}
			\put(101,-12){$\text{(d)}$}
		\end{overpic}
	\end{minipage}
	\begin{minipage}{0.49\linewidth}
	\vspace{0.5cm}
	\begin{overpic}[width=1\linewidth]{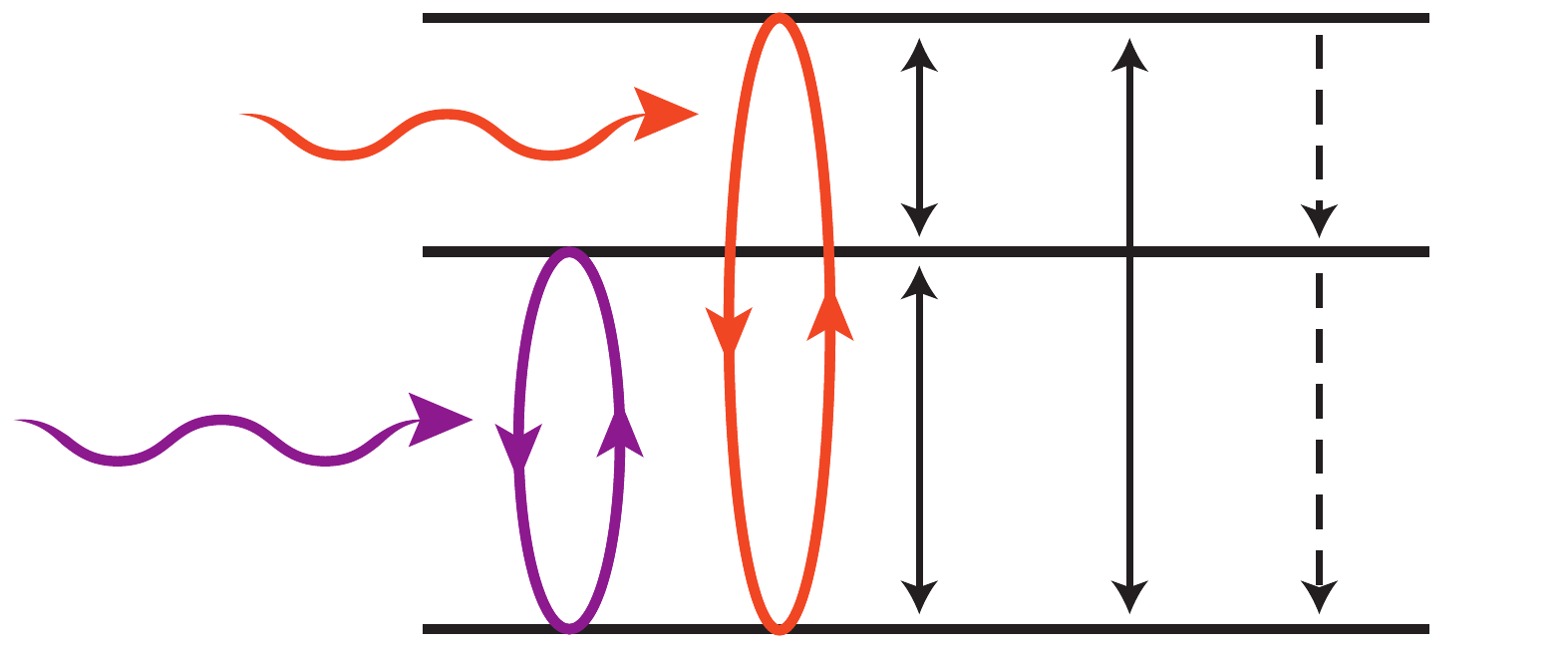}
		\put(6,18){ $\Omega_{p}$}
		\put(6, 33){ $\Omega_{d}$}
		\put(91, 38){ $|2 \rangle$}
		\put(91,24){ $|1 \rangle$}
		\put(91, 0){ $|0 \rangle$}
		\put(60,12){$\omega_{10}$}
		\put(60,32){$\omega_{21}$}
		\put(74,12){$\omega_{20}$}
		\put(86,32){$\Gamma_{21}$}
		\put(86,12){$\Gamma_{10}$}
	\end{overpic}
\end{minipage}
	\begin{minipage}{0.45\linewidth}
		\begin{overpic}[width=1\linewidth]{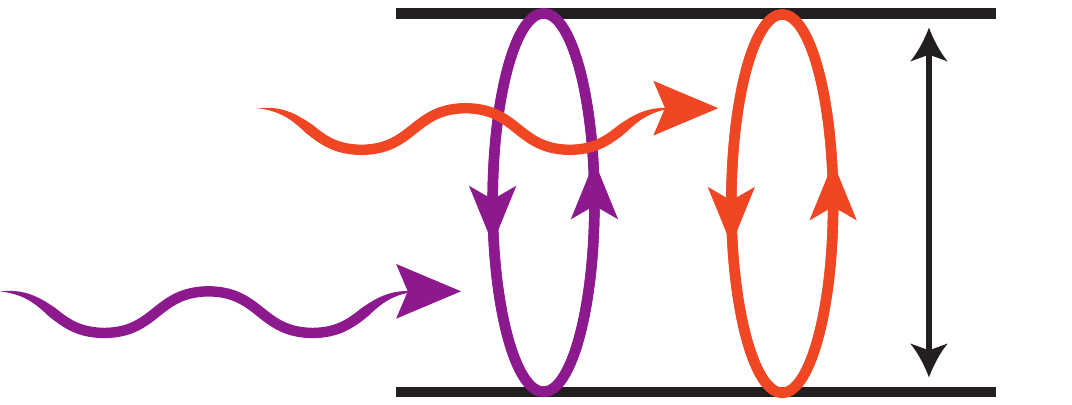}
			\put(15,26){ $\Omega_{d}$}
			\put(15, 14){ $\Omega_{p}$}
			\put(89,17){$\omega_{10}$}
			\put(92, 34){ $|1 \rangle$}
			\put(92, 0){ $|0 \rangle$}
		\end{overpic}
	\end{minipage}
	\begin{minipage}{0.53\linewidth}
		\vspace{0.5cm}
	\begin{overpic}[width=1\linewidth]{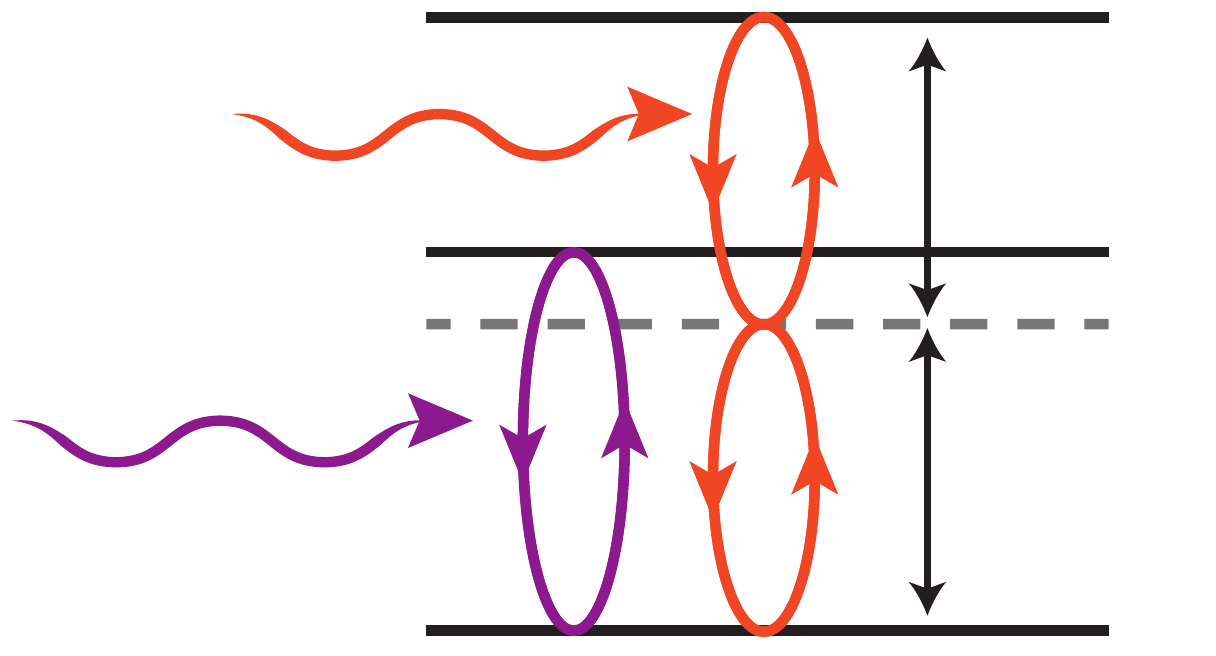}
		\put(10,23){ $\Omega_{p}$}
		\put(10, 43){ $\Omega_{d}$}
		\put(90, 50){ $|2 \rangle$}
		\put(90,32){ $|1 \rangle$}
		\put(90, 1){ $|0 \rangle$}
		\put(80,40){$\frac{\omega_{20}}{2}$}
		\put(80,13){$\frac{\omega_{20}}{2}$}
	\end{overpic}
\end{minipage}
	\caption{Sketches of the systems considered in this article.
	(a) All systems we study are variations on a setup with an atom in front of a mirror (short). The atom interacts with a strong drive field (orange) and a weak probe field (purple).
	(b) A three-level atom with transition frequencies $\omega_{ij}$ between the states $\ket{i}$ and $\ket{j}$. The atom is strongly driven with a drive amplitude $\Omega_d$ on the $\ket{0} \leftrightarrow \ket{2}$ transition. The probe field with amplitude $\Omega_p$ is applied to the $\ket{0} \leftrightarrow \ket{1}$ transition. The decay rates are denoted by $\Gamma_{ij}$ for the transition from $\ket{i}$ to $\ket{j}$. 
	(c) A two-level system both driven and probed around its transition frequency $\omega_{10}$. 
	(d) A three-level atom like in (b), but driven at half the frequency of the $\ket{0} \leftrightarrow \ket{2}$ transition, resulting in a two-photon driving that dresses all states of the weakly anharmonic system.}
	\label{fig:System}
\end{figure}
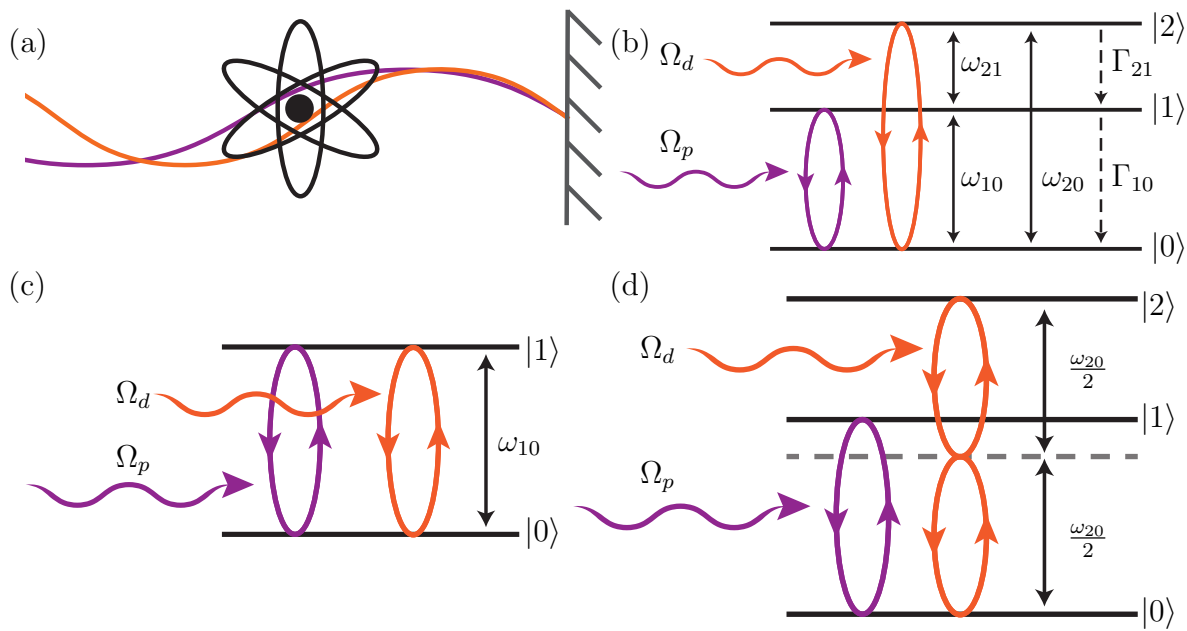
%===================================================

We study all three amplification mechanisms outlined above for an atom, with either two or three levels, coupled to a 1D waveguide terminated by a mirror (in the form of a short) at one end, as depicted in \figref{fig:System}. We show that this setup has two advantages over the corresponding one in an open waveguide, leading to a doubling or more of the maximum amplification that we can achieve. Firstly, the mirror reflects the electromagnetic field such that we only have one input-output channel, which avoids losing half of the atomic output in one direction, as happens in an open waveguide. The second advantage of the mirror setup is that, due to interference effects, the coupling to the waveguide is set by the position of the atom and its transition frequency. This enables manipulation of the relative coupling strengths for different transitions in a three-level atom, either by changing the atomic frequency, which is possible in superconducting circuits~\cite{You2011, Gu2017, Kockum2019a, Blais2020}, or by changing the distance from the atom to the mirror. We note that the coupling strengths can also be made frequency-dependent and tunable using giant atoms~\cite{Kockum2020, Kockum2014, Gustafsson2014, Guo2017, Manenti2017, Kockum2018, Karg2019, Ask2019, Gonzalez-Tudela2019, Guimond2020, Guo2020, Kannan2020, Vadiraj2020, Wang2020, Ask2020}, which couple to the waveguide at multiple points, but if the giant atom is placed in an open waveguide, the problem of losing half the output remains. With a mirror, a single small atom is simpler to implement than a giant atom, but still sufficient to achieve the advantageous frequency-dependent coupling.

The first system we consider is shown in \figpanel{fig:System}{b} and discussed in \secref{sec:3Levels1Photon}. It is a three-level atom with a strong drive on the transition between the ground state $\ket{0}$ and the second excited state $\ket{2}$. When the decay rate from $\ket{2}$ to the first excited state $\ket{1}$ is larger than the decay rate from $\ket{1}$ to $\ket{0}$, a population inversion between $\ket{1}$ and $\ket{0}$ is created. This leads to amplification of a weak probe signal on the $\ket{0} \leftrightarrow \ket{1}$ transition. We find that with the mirror, a maximum amplitude gain of \unit[25]{\%} can be reached, whereas the maximum amplification in an open waveguide is \unit[12.5]{\%}~\cite{Astafiev2010}.

Next, we study, in \secref{sec:TwoLevelStrong}, the resonantly driven two-level atom depicted in \figpanel{fig:System}{c}. The strong drive splits the energy levels of the atom and enable transitions in the dressed-state basis. By probing the system in the vicinity of the bare resonance frequency, we achieve a maximal amplitude gain of around \unit[6.9]{\%} with the mirror. For an open waveguide, we find an amplification of around \unit[3.4]{\%} for the same system parameters. In contrast to the previous case, amplification is not due to population inversion, but enabled by higher-order processes between the dressed states~\cite{Friedmann1987, Wen2018}.

The last system we study, in \secref{sec:3Levels2Photon}, is the three-level system shown in \figpanel{fig:System}{d}, driven at half the $\ket{0} \leftrightarrow \ket{2}$ transition frequency. Similarly to the strongly driven two-level system, the energy levels are split by the driving and transitions take place between dressed states. By probing the system, we find a maximal amplification of around \unit[6.2]{\%} with a mirror, which exceeds the amplification of the same setup in an open waveguide by more than a factor 2~\cite{Koshino2013}. Here the amplification is due to hidden inversion --- population inversion between the dressed states of the system.

We further show that all these systems can be realized with currently available state-of-the-art technology in experimental waveguide-QED setups with a transmon qubit~\cite{Koch2007} coupled to a 1D transmission line. The transition frequencies of such a qubit are tunable in situ, which means that the ratio of the decay rates of the transmon can be chosen to reach the optimal amplification settings. Our proposed setups, which represent an ultimate quantum limit for amplification, may thus find applications in superconducting quantum information processing.

%%%%%%%%%%%%%%%%%%%%%%%%%%%%%%%%%%%%%%%%%%%%%%%

\section{Amplification with a strongly driven three-level atom in front of a mirror}
\label{sec:3Levels1Photon}

We begin by studying the setup with a three-level atom in front of a mirror shown in \figpanel{fig:System}{b}. The system is coherently driven with amplitude $\Omega_d$ on the transition between the ground state $\ket{0}$ and the second excited state $\ket{2}$. The aim is to create a population inversion between the states $\ket{0}$ and $\ket{1}$, which can lead to a gain in the reflection of a weak coherent probe resonant with the $\ket{0} \leftrightarrow \ket{1}$ transition. To achieve population inversion, the life-time $1/\Gamma_{10}$ of the first excited state should be much longer than the life-time $1/\Gamma_{21}$ of the second excited state. Here, we assume that the atom is a good approximation of a ladder-type $\Xi$ system with $\Gamma_{20} \ll \Gamma_{10}, \Gamma_{21}$.

%%%%%%%%%%%%%%%%%%%%%%%%%%

\subsection{Hamiltonian and master equation}

The Hamiltonian of the system in the frame rotating at the drive frequencies is (we set $\hbar = 1$ throughout this article)
\bea
H &= H_a + H_{\rm int}, \\
H_a &= \delta \omega_{10} \sigma_{11} + \delta \omega_{20} \sigma_{22}, \\
H_{\text{int}} &= \frac{\Omega_d}{2} \mleft( \sigma_{20} + \sigma_{02} \mright) + \frac{\Omega_p}{2} \mleft( \sigma_{10} + \sigma_{01} \mright).
\eea
where $\sigma_{ij} = \ketbra{i}{j}$, the drive amplitude on the $\ket{0} \leftrightarrow \ket{2}$ ($\ket{0} \leftrightarrow \ket{1}$) transition is given by $\Omega_d$ ($\Omega_p$), and $\delta \omega_{ij} = \omega_{ij} - \omega_{ij}^d$ for $i>j$ is the detuning between the transition frequency $\omega_{ij} = \omega_i - \omega_j$ and the frequency $\omega_{ij}^d$ of the drive on that transition. The dynamics of the system is described by the master equation
\be
\dot \rho = -\frac{i}{\hbar} \comm{H}{\rho} + \mathcal{L} \mleft[ \rho \mright]
\label{eq:me}
\ee
for the density matrix $\rho = \sum_{i,j} \rho_{ij} \ketbra{i}{j}$. The Lindbladian term in \eqref{eq:me} is given by
\be
\mathcal{L} \mleft[ \rho \mright] = \Gamma_{21} \rho_{22} \mleft( - \sigma_{22} + \sigma_{11} \mright) + \Gamma_{10} \rho_{11} \mleft( - \sigma_{11} + \sigma_{00} \mright) - \sum_{i \neq j} \gamma_{ij} \rho_{ij} \sigma_{ij},
\ee
with the dephasing $\gamma_{ij} = \gamma_{ji}$ and the relaxation rates $\Gamma_{ij}$ between the states $\ket{i}$ and $\ket{j}$, $i > j$. Since we assume negligible temperature, we can neglect thermal excitations, i.e., $\Gamma_{01} = \Gamma_{12} = \Gamma_{02} = 0$.

%%%%%%%%%%%%%%%%%%%%%%%%%%

\subsection{Steady-state solution}

We assume that the probe on the $\ket{0} \leftrightarrow \ket{1}$ is weak, i.e., $\Omega_{10}/ \Gamma_{10} \ll 1$. Solving the master equation for the steady state ($\dot \rho = 0$), we obtain~\cite{Astafiev2010}
\bea
\rho_{00} %&= \frac{\frac{2 \Gamma_{32} \left( \gamma_{31}^2 + \delta \omega_{31}^2 \right)}{\gamma_{31} \Omega_{31}^2} + 1}{\frac{2 \Gamma_{23} \left( \gamma_{31}^2 + \delta \omega_{31}^2 \right)}{\gamma_{31} \Omega_{31}^2} + \frac{\Gamma_{32}}{\Gamma_{21}} + 2} \\
%&= \frac{\frac{2 \Gamma_{32} \left| \lambda_{13} \right|^2}{\gamma_{31} \Omega_{31}^2} + 1}{\frac{2 \Gamma_{23}\left| \lambda_{13} \right|^2}{\gamma_{31} \Omega_{31}^2} + \frac{\Gamma_{32}}{\Gamma_{21}} + 2} \\
&=& \frac{A}{A+B+1}, \quad
\rho_{11} %&=  \frac{\frac{\Gamma_{32}}{\Gamma_{21}}}{\frac{2 \Gamma_{23} \left( \gamma_{31}^2 + \delta \omega_{31}^2 \right)}{\gamma_{31} \Omega_{31}^2} + \frac{\Gamma_{32}}{\Gamma_{21}} + 2} \\
%&= \frac{\frac{\Gamma_{32}}{\Gamma_{21}}}{\frac{2 \Gamma_{23}\left| \lambda_{13} \right|^2}{\gamma_{31} \Omega_{31}^2} + \frac{\Gamma_{32}}{\Gamma_{21}} + 2} \\
= \frac{B}{A+B+1}, \quad
\rho_{22} %&= \frac{1}{\frac{2\Gamma_{23}\left| \lambda_{13} \right|^2}{\gamma_{31} \Omega_{31}^2} + \frac{\Gamma_{32}}{\Gamma_{21}} + 2} \\
= \frac{1}{A+B+1}, \label{eq:rho001122} \\
\rho_{10} &=& \frac{i \frac{\Omega_p}{2 \lambda_{10}} \mleft( \frac{\Omega_d^2}{4 \lambda_{02} \lambda_{12}} \mleft( \rho_{00} - \rho_{22} \mright) + \rho_{11} - \rho_{00} \mright)}{1 + \frac{\Omega_d^2}{4 \lambda_{10} \lambda_{12}}},
\eea
where
\be
A = \frac{\rho_{00}}{\rho_{22}} = \frac{2 \Gamma_{21} \abssq{\lambda_{02}}}{\gamma_{20} \Omega_d^2} + 1, \quad B = \frac{\rho_{11}}{\rho_{22}} = \frac{\Gamma_{21}}{\Gamma_{10}},
\label{eq:AB}
\ee
and $\lambda_{i j} = \lambda_{j i}^{*}$ with
\be
\lambda_{10} = \gamma_{10} + i \delta \omega_{10}, \quad
\lambda_{12} = \gamma_{21} - i \delta \omega_{20} + i \delta \omega_{10}, \quad
\lambda_{02} = \gamma_{20} - i \delta \omega_{20}.
\ee
%

%%%%%%%%%%%%%%%%%%%%%%%%%%

\subsection{Amplification and optimal drive strength}

We now assume that $\Gamma_{21} \gg \Gamma_{10}$ to ensure population inversion. From Eqs.~(\ref{eq:rho001122}) and (\ref{eq:AB}), we see that the second excited state then is nearly unpopulated. Neglecting terms of order $\mathcal{O} (\Gamma_{10} / \Gamma_{21})$, this gives
\be
\rho_{10} \approx i \frac{\Omega_p \mleft( \rho_{11} - \rho_{00} \mright)}{2 \lambda_{10} + \frac{\Omega_d^2}{2 \lambda_{12}}}.
\ee
The reflection coefficient is given by~\cite{Hoi2015, Lu2020}
\be
r = 1 - 2i \frac{\Gamma_{10}}{\Omega_p} \expec{\sigma_{01}} = 1 - 2i \frac{\Gamma_{10}}{\Omega_p} \rho_{10} = 1 + 2 \Gamma_{10} \frac{ \mleft( \rho_{11} - \rho_{00} \mright)}{2 \lambda_{10} + \frac{\Omega_d^2}{2 \lambda_{12}}}.
\label{eq:r}
\ee
This is where our derivation deviates from that in Ref.~\cite{Astafiev2010} for an open waveguide. The mirror adds a factor 2 to the second term of the reflection coefficient compared to the open waveguide.

From \eqref{eq:r}, it is clear that amplification requires $\rho_{11} > \rho_{00}$, which leads to
\be
\frac{\Gamma_{21}}{\Gamma_{10}} > \frac{2 \Gamma_{21} \abssq{\lambda_{02}}}{\gamma_{20} \Omega_d^2} + 1
\Rightarrow \Omega_d^2 > \frac{2 \Gamma_{10} \abssq{\lambda_{02}}}{\gamma_{20}} + \frac{\Omega_d^2 \Gamma_{10}}{\Gamma_{21}} \approx \frac{2 \Gamma_{10} \abssq{\lambda_{02}}}{\gamma_{20}}.
\ee
For a resonant drive, $\delta \omega_{20} = 0$, this reduces to
\be
\Omega_d^2 > 2 \Gamma_{10} \gamma_{20}
\ee
If we further assume no pure dephasing, we have $\gamma_{20} = \Gamma_{21}/2$ and thus
\be
\Omega_d^2 > \Gamma_{10} \Gamma_{21}.
\ee

We now calculate the maximal possible amplitude gain of the single-atom amplifier. We consider double resonance, $\delta \omega_{20} = \delta \omega_{10} = 0$, and no pure dephasing, i.e., $\gamma_{10} = \Gamma_{10}/2$, $\gamma_{21} = \Gamma_{21}/2$, and $\gamma_{20} = \Gamma_{21}/2$. We can then rewrite the populations in \eqref{eq:rho001122} as $\rho_{00} = 1/\left(1 + \nu \right)$ and $\rho_{11} = \nu / \mleft( 1 + \nu \mright)$ with $\nu = \Omega_d^2 / \mleft(\Gamma_{10} \Gamma_{21} \mright)$, which leads to the reflection coefficient
\be
r = 1 + 2 \frac{\mleft( \nu - 1 \mright)}{\mleft(1 + \nu \mright)^2}.
\ee
This expression reaches its maximum value for $\nu = 3$, which is achieved when
\be
\Omega_d^2 = 3 \Gamma_{10} \Gamma_{21}.
\ee
With this value for the drive amplitude, the maximum reflection is given by
\be
	\abs{r} = 1 + \frac{1}{4},
\ee
which corresponds to an amplitude gain of \unit[25]{\%}. If we include higher orders of $\Gamma_{10}/\Gamma_{21}$ in the calculation, the first-order correction to the maximum value of the reflection becomes
\be
r = 1 + \frac{1}{4} - \frac{3}{8} \frac{\Gamma_{10}}{\Gamma_{21}} + \mathcal{O} \mleft( \frac{\Gamma_{10}}{\Gamma_{21}} \mright)^2.
\ee

%===================================================
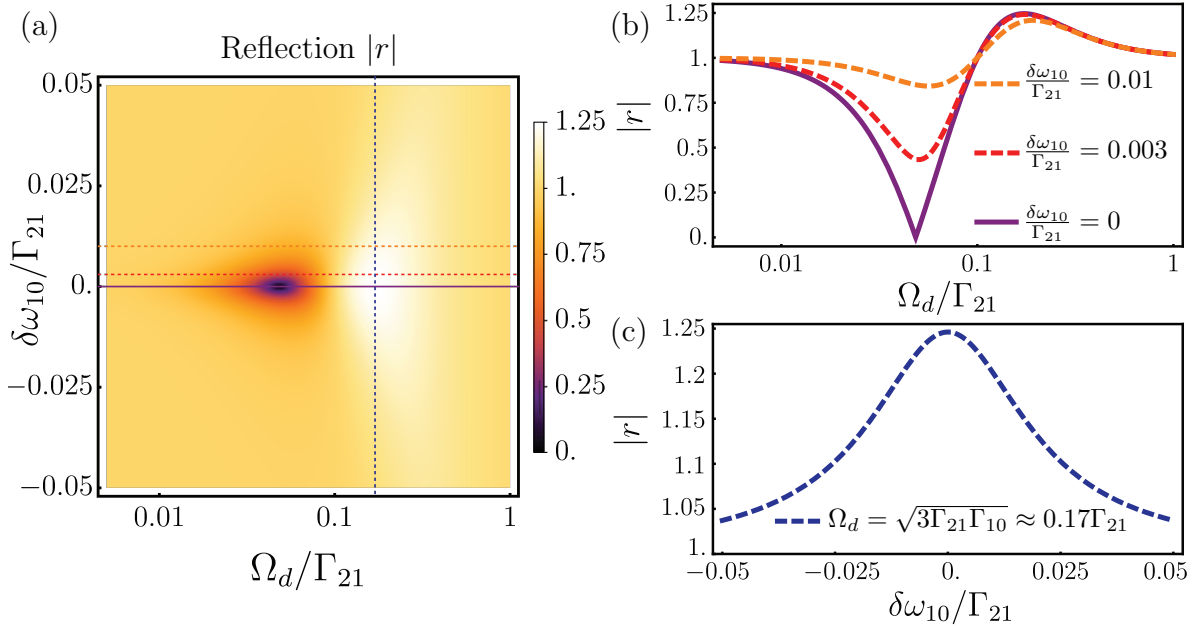
\begin{figure}[t]
	\centering
	\begin{overpic}[width=\textwidth]{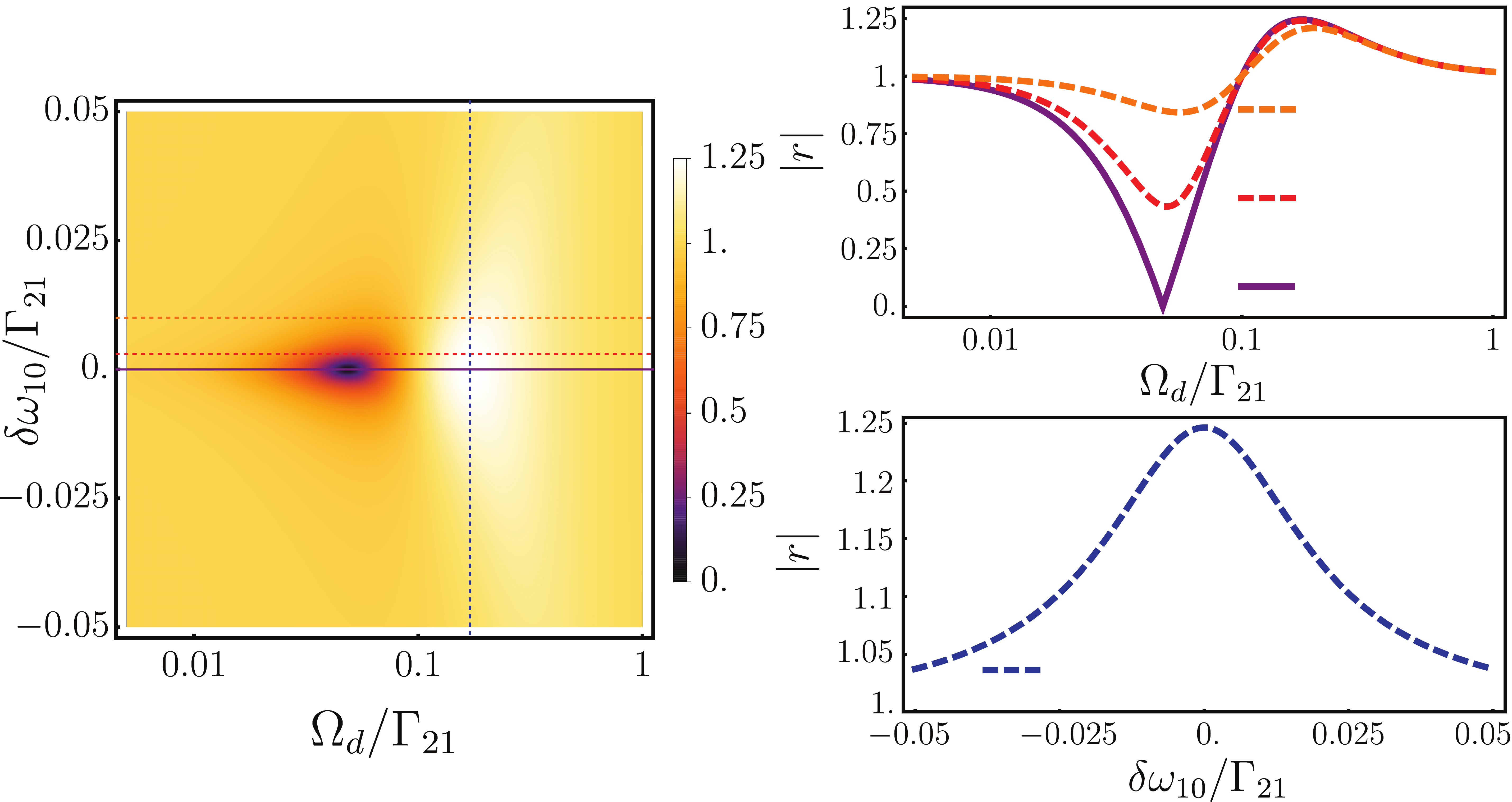}%Summary/WeaklyDriven/WeaklyDriven.nb
	        \put(18,48){Reflection $|r|$}
			\put(1,50){$\text{(a)}$}
			\put(51,50){$\text{(b)}$}
			\put(51,24){$\text{(c)}$}
			\put(86,45.5){\footnotesize $\frac{\delta \omega_{10}}{ \Gamma_{21}}= 0.01$}
		    \put(86,39.5){\footnotesize $\frac{\delta \omega_{10}}{ \Gamma_{21}}= 0.003$}
		    \put(86,33.5){\footnotesize $\frac{\delta \omega_{10}}{ \Gamma_{21}}= 0$}
		    \put(69.5,8.2){\footnotesize $\Omega_d = \sqrt{3 \Gamma_{21} \Gamma_{10}} \approx 0.17 \Gamma_{21}$}
	\end{overpic}
	\caption{Optimizing the amplitude gain of a strongly driven three-level atom in front of a mirror.
	(a) The absolute value of the reflection coefficient, $\abs{r}$, i.e., the amplitude gain, as a function of the drive strength $\Omega_d / \Gamma_{21}$ and the probe detuning $\delta \omega_{10} / \Gamma_{21}$. The reflection is calculated for a resonant drive, $\delta \omega_{20} = 0$, no pure dephasing, $\gamma_{21} = \gamma_{20} = \Gamma_{21}/2$ and $\gamma_{10} = \Gamma_{10}/2$, and with $\Gamma_{10}/\Gamma_{21} = 0.01$. The reflection reaches its maximum at $\Omega_{20} = \sqrt{3 \Gamma_{21} \Gamma_{10}} \approx 0.17 \Gamma_{21}$.
	(b) Horizontal linecuts from panel (a) showing the amplitude gain as a function of $\Omega_d$ for three different values of the probe detuning: $\delta \omega_{10} = 0$ (purple, solid curve), $\delta \omega_{10} / \Gamma_{21}= 0.003$ (red, dashed), and $\delta \omega_{10} / \Gamma_{21}= 0.01$ (orange, dashed). The maximum values for the orange and red curves are slightly below $1.25$.
	(c) Vertical blue dashed linecut from panel (a) showing the reflection at the optimal driving strength $\Omega_d = \sqrt{3 \Gamma_{21} \Gamma_{10}} \approx 0.17 \Gamma_{21}$ as a function of the detuning $\delta \omega_{10} / \Gamma_{21}$.}
	\label{fig:ReflWeak}
\end{figure}

In \figpanel{fig:ReflWeak}{a}, we plot the absolute value of the reflection coefficient as a function of the drive amplitude $\Omega_d$ and the detuning $\delta \omega_{10}$. In \figpanel{fig:ReflWeak}{b}, we further illustrate the effect of non-zero detuning with a few linecuts from \figpanel{fig:ReflWeak}{a}. It is clear that the maximum reflection is achieved on resonance.

%%%%%%%%%%%%%%%%%%%%%%%%%%

\subsection{Optimal population inversion}

In the previous subsection, we found that the drive strength $\Omega_d^2 = 3 \Gamma_{10} \Gamma_{21}$ gives the highest amplitude gain. Inserting this into the expressions for $\rho_{00}$ and $\rho_{11}$ in \eqref{eq:rho001122}, we obtain
\be
\rho_{00} = \frac{1}{\frac{4}{3} + 2 \underbrace{\frac{\Gamma_{10}}{\Gamma_{21}}}_{\ll 1}} \approx \frac{3}{4} , \quad \rho_{11} = \frac{\frac{1}{3} + \frac{\Gamma_{10}}{\Gamma_{21}}}{\frac{4}{3} + 2 \frac{\Gamma_{10}}{\Gamma_{21}}} \approx \frac{1}{4}.
\ee
We see that in order to maximize the amplitude gain, the population is not completely inverted (that would be $\rho_{11} = 1$, $\rho_{00} = 0$). For a complete population inversion to happen, the following condition has to be fulfilled:
\be
\frac{B}{A} = \frac{\rho_{11}}{\rho_{00}} \gg 1 \quad \Rightarrow \quad \frac{\Gamma_{21}}{\Gamma_{10}} \gg \frac{2 \Gamma_{21} \gamma_{20}}{\Omega_d^2}.
\ee
This could be achieved by further increasing the drive strength $\Omega_d$. However, if we look at the expression for the reflection in \eqref{eq:r}, we see that increasing the pumping strength towards infinity would make the reflection revert to 1. This trade-off explains why we do not achieve a maximum amplitude gain of $\sqrt{2}$ (a power gain of 2), which would be the result if an incoming photon would stimulate emission of another photon from a perfectly inverted atom.

%%%%%%%%%%%%%%%%%%%%%%%%%%

\subsection{Correction with pure dephasing}

Now we discuss the effect of pure dephasing on the previous results. Neglecting terms of order $\mathcal{O} (\Gamma_{10} / \Gamma_{21})$, the reflection coefficient on resonance with pure dephasing included can be written as
\be
r = 1 + 2 \frac{\Gamma_{10}}{\gamma_{10}}\frac{\mleft( \eta - 1 \mright)}{\mleft( \eta + 1 \mright) \mleft( \eta \frac{\Gamma_{10} \gamma_{20}}{\gamma_{21} \gamma_{10}} + 2 \mright)}
\ee
with $\eta = \frac{\Omega_d^2}{2 \Gamma_{10} \gamma_{20}}$. Maximizing this expression, we find the optimal value for $\eta$:
\be
\eta_{\rm max} = 1 + \eta_c, \quad \eta_c = \sqrt{2 \mleft( 1 + 2 \frac{\gamma_{21} \gamma_{10}}{\Gamma_{10} \gamma_{20}}\mright) }
%\eta_c = \frac{\sqrt{2 \frac{\Gamma_{10}}{\gamma_{10}} \gamma_{20} \mleft( \frac{\Gamma_{10}}{\gamma_{10}} \gamma_{20} + 2 \gamma_{21} \mright)}}{ \frac{\Gamma_{10}}{\gamma_{10}} \gamma_{20}}.
\ee
Hence, the optimal drive strength including pure dephasing is
\be
\Omega_d^2 = 2 \Gamma_{10} \gamma_{20} \mleft( 1 + \eta_c \mright),
\ee
for which we obtain the maximum reflection
\be
r = 1 + 2 \frac{\Gamma_{10}}{\gamma_{10}} \frac{1}{ \mleft[ 1 + \sqrt{2} \mleft( 1 + 2 \frac{\gamma_{21} \gamma_{10}}{\Gamma_{10} \gamma_{20}} \mright)^{-1/2}  \mright]  \mleft[ \mleft( \sqrt{2 \mleft( 1 + 2 \frac{\gamma_{21} \gamma_{10}}{\Gamma_{10} \gamma_{20}}\mright) }  +  1    \mright) \frac{\Gamma_{10} \gamma_{20}}{\gamma_{21} \gamma_{10}} +2 \mright] }.
\ee
%

%%%%%%%%%%%%%%%%%%%%%%%%%%

\subsection{Experimental feasibility}

We now apply the theoretical results above to a typical experimental system of a superconducting transmon qubit~\cite{Koch2007} to see what the optimal parameters for an experiment would be, and whether they are within reach for currently available devices. By shorting one end of the transmission line to create an effective mirror, the decay rates $\Gamma_{10}$ and $\Gamma_{21}$ become a function of the transition frequency $\omega_{10}$ of the energy levels~\cite{Hoi2015}
\bea
\Gamma_{10} =&\  2 \Gamma_{10}^{\text{TL}} \cos^2 \mleft[ \frac{L}{v} \omega_{10} \mright], \label{eq:Gamma10}\\
\Gamma_{21} =&\  2 \Gamma_{21}^{\text{TL}} \cos^2 \mleft[ \frac{L}{v} \mleft( \omega_{10} + \alpha \mright) \mright], \label{eq:Gamma21}
\eea
where $\Gamma_{10}^{\text{TL}}/2\pi = \unit[37.5]{MHz}$ and $\Gamma_{21}^{\text{TL}} /2\pi \approx 2 \Gamma_{10}^{\text{TL}}/2\pi = \unit[75]{MHz}$ are the bare relaxation rates in an open transmission line, $\alpha = \omega_{21} - \omega_{10}$ is the anharmonicity between the transition frequencies, $L = \unit[33]{mm}$ is the distance between the transmon and the mirror, and $v = \unit[9 \cdot 10^7]{m/s}$ is the speed of light in the transmission line. The given values are typical for this kind of setup~\cite{Hoi2015, Wen2018}.

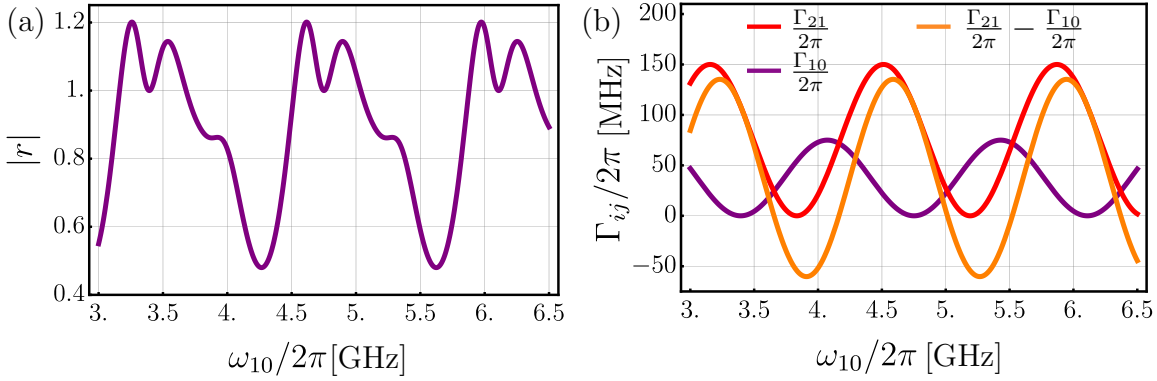
\begin{figure}[t]
	\centering
	\begin{minipage}{0.48\linewidth}
		\begin{overpic}[width=\textwidth]{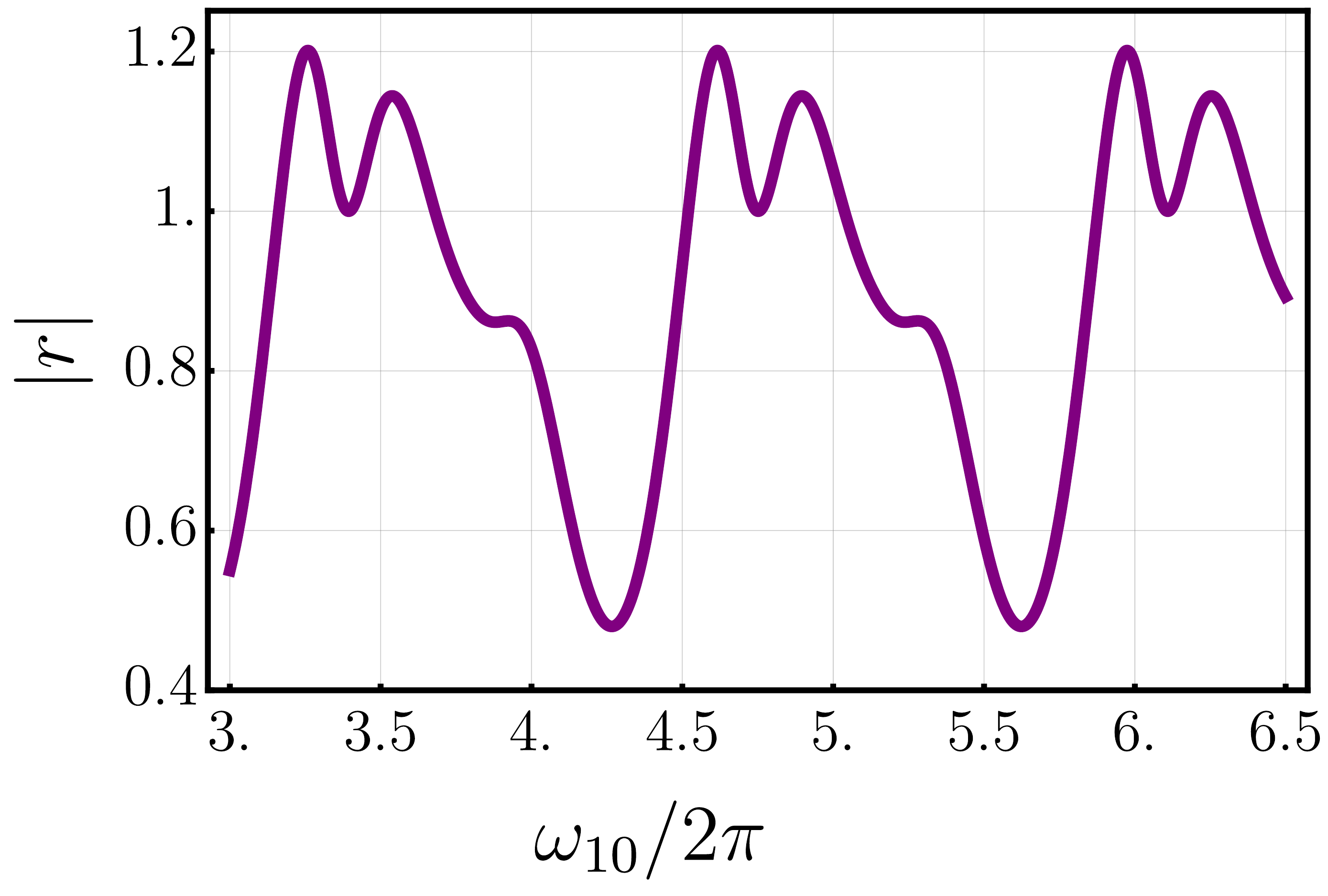}%Summary/WeaklyDriven/WeaklyDriven.nb
			\put(1,62){$\text{(a)}$}
			\put(102.5,62){$\text{(b)}$}
			\put(58,2.5){[\unit[]{GHz}]}
		\end{overpic}
	\end{minipage}
	\begin{minipage}{0.49\linewidth}
		\begin{overpic}[width=\textwidth]{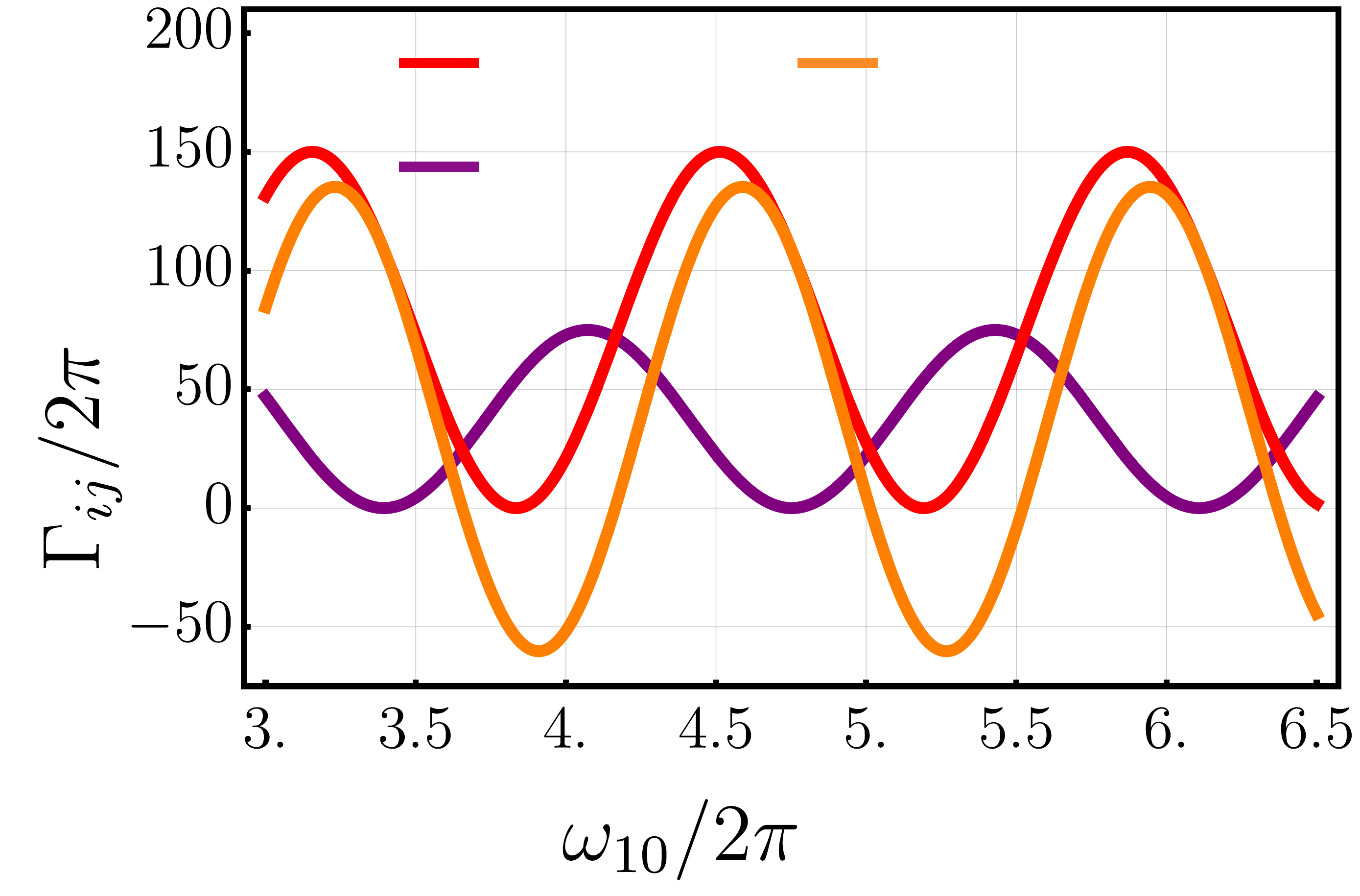}%Summary/WeaklyDriven/WeaklyDriven.nb
			\put(35,52){ $\frac{\Gamma_{10}}{2\pi}$}
			\put(35,59.5){ $\frac{\Gamma_{21}}{2\pi}$}
			\put(65,59.5){ $\frac{\Gamma_{21}}{2\pi} - \frac{\Gamma_{10}}{2\pi}$}
			\put(60,2.5){[\unit[]{GHz}]}
			\put(3.5,42){\rotatebox{90}{[\unit[]{MHz}]}}
		\end{overpic}
	\end{minipage}
	\caption{Reflection coefficient and decay rates as functions of transition frequency.
	(a) Absolute value of the reflection coefficient as a function of transition frequency $\omega_{10}$ at drive strength $\Omega_d / 2\pi= \unit[59.5]{MHz}$ and pure dephasing $\Gamma_{10}^\phi / 2 \pi = \unit[1.65]{MHz}$, $\Gamma_{21}^\phi /2\pi = \Gamma_{20}^\phi /2\pi = \unit[5]{MHz}$.
	(b) The decay rates $\Gamma_{10} / 2 \pi$ (purple) and $\Gamma_{21} / 2 \pi$ (red), and their difference $\Gamma_{21} / 2 \pi - \Gamma_{10} / 2\pi$ (orange), as a function of the transition frequency $\omega_{10}$.
	\label{fig:ReflMax}}
\end{figure}

The transition frequencies of the transmon are tunable in situ by an external magnetic flux, so we want to find the resonance frequency $\omega_{10}$ that gives the highest possible reflection. We therefore express the reflection as a function of the drive strength $\Omega_d$ and the frequency $\omega_{10}$, using Eqs.~(\ref{eq:Gamma10})--(\ref{eq:Gamma21}), and maximize this function numerically. A plot of the resulting reflection amplitude can be seen in \figpanel{fig:ReflMax}{a}. With the above parameters and pure dephasing rates of $\Gamma_{10}^\phi /2\pi = \unit[1.65]{MHz}$, $\Gamma_{21}^\phi/2\pi = \Gamma_{20}^\phi = \unit[5]{MHz}$, again chosen from typical values~\cite{Hoi2015, Wen2018, Vadiraj2020} and optimal drive strength $\Omega_d / 2 \pi = \unit[59.5]{MHz}$, the reflection reaches a maximum of $1.2$ which corresponds to an amplitude gain of \unit[20]{\%}. Due to the non-zero dephasing, this is lower than the theoretical limit of \unit[25]{\%} calculated above. We note that dephasing and non-radiative decay rates can be lower than what we have assumed here, as shown, e.g., in Refs.~\cite{Mirhosseini2019, Scigliuzzo2020}.

We find that there are two local maxima for the gain in \figpanel{fig:ReflMax}{a}, located in the area close to the nodes of the decay rate $\Gamma_{10}$ [see \figpanel{fig:ReflMax}{b}], e.g., between $\omega_{10} / 2 \pi \approx \unit[4.5]{GHz}$ and $\omega_{10} / 2 \pi \approx \unit[5.0]{GHz}$. This is the area where the requirement for amplification $\Omega_d^2 > 3 \Gamma_{10} \Gamma_{21}$ is fulfilled. Between the two local maxima, we find a local minimum with a \unit[0]{\%} gain. This local minimum occurs at the node of the electromagnetic field, where the decay rate $\Gamma_{21}$ goes to zero and no population inversion is possible [see \figpanel{fig:ReflMax}{b}].

%===================================================
%===================================================

We note that if tuning the qubit frequency to the maxima in \figpanel{fig:ReflMax}{a} is required, the amplifier will be limited to working at these frequencies, with rather narrow bandwidth. However, this bandwidth can be increased by making the mirror itself tunable, e.g., by placing a superconducting quantum interference device (SQUID) at the end of the waveguide~\cite{Sandberg2008, Johansson2009, Wilson2011}. In this way, the interference can be changed such that the maxima are moved to other qubit frequencies.

We also note that driving the $\ket{0} \leftrightarrow \ket{2}$ transition directly is hard in a transmon due to selection rules~\cite{Koch2007}. It is possible to drive the transition with a two-photon drive instead, where the frequencies of two drive photons sum up to $\omega_{20}$. However, if this drive is too strong, the qubit states will be dressed and we will instead have a setup like that discussed in \secref{sec:3Levels2Photon}. Another solution is to use another superconducting qubit which does not suffer from this limitation on allowed transitions, e.g., the flux qubit, as was done in Ref.~\cite{Astafiev2010}.

Finally, if the $\ket{0} \leftrightarrow \ket{2}$ transition is driven through the waveguide, the strength with which the drive couples to the system will be frequency-dependent in the same way as in \eqref{eq:Gamma10}, with $\omega_{10}$ replaced by $\omega_{20}$. As long as the drive frequency does not correspond to a node of the field at the atom, a decrease or increase in coupling strength can be compensated by adjusting the input drive power. It would also be possible to avoid any such issues by driving the atom through a separate line not affected by the interference with the mirror.

%%%%%%%%%%%%%%%%%%%%%%%%%%%%%%%%%%%%%%%%%%%%%%%

\section{Amplification with a strongly driven two-level atom in front of a mirror}
\label{sec:TwoLevelStrong}

The next setup we consider is a two-level atom in front of a mirror that is driven strongly on resonance, as depicted in \figpanel{fig:System}{c}. The strong driving results in a splitting of the atomic energy levels such that the dynamics are best understood in terms of dressed states. As shown theoretically in Refs.~\cite{Mollow1972, Friedmann1987} and experimentally in Refs.~\cite{Wu1977, Xu2007, Wen2018}, amplification can be achieved in this setup through higher-order processes when probing at frequencies in-between those of the Mollow triplet.

%%%%%%%%%%%%%%%%%%%%%%%%%%

\subsection{Hamiltonian and equations of motion}
\label{subsec:TwoLevelStrong_HEOM}

The Hamiltonian of a driven two-level system, with ground state $\ket{0}$ and excited state $\ket{1}$, interacting with the continuum of modes in the semi-infinite waveguide, is, in a frame rotating with the drive frequency $\omega_d$,
\bea
H =&\ H_a + H_f + H_\text{int}, \label{eq:Hstrong}\\
H_a =&\ \delta \omega_{10} \sigma_{11} + E \mleft( \sigma_t + \sigma_t^\dag \mright), \label{eq:Ha2strong}\\
H_f =&\ \int d \omega \ \omega a^\dag (\omega) a (\omega),  \label{eq:Hfstrong}\\
H_\text{int} =&\ \frac{1}{\sqrt{2 \pi}} \int_0^\infty d \omega \mleft( a^\dag (\omega) \sigma_t + \sigma_t^\dag a (\omega) \mright), \label{eq:Hafstrong}
\eea
where $\delta \omega_{10} = \omega_{10} - \omega_d$, $\Omega_d = 2\sqrt{\Gamma_{10}} E$, $\abssq{E}$ is the number of incoming drive photons per second, $\sigma_t = \sqrt{\Gamma_{10}} \sigma_{01}$, and $a^\dag (\omega)$ [$a (\omega)$] are the photon creation [annihilation] operators at frequency $\omega$.
We calculate the eigenenergies $\omega_g$, $\omega_e$ and the corresponding dressed eigenstates $\ket{g}$, $\ket{e}$ of the atomic Hamiltonian $H_a$ in \eqref{eq:Ha2strong}, and define the population and transition operators for the dressed states as
\bea
\sigma_{\mu \nu} &= \ketbra{\mu}{\nu},
%	\sigma_{mm} &= \ket{m} \bra{m} \\
%	\sigma_{ee} &= \ket{e} \bra{e} \\
%	\sigma_{gm} &= \ket{g} \bra{m} \\
%	\sigma_{mg} &= \ket{m} \bra{g} \\
%	\sigma_{ge} &= \ket{g} \bra{e} \\
%	\sigma_{eg} &= \ket{e} \bra{g} \\
%	\sigma_{me} &= \ket{m} \bra{e} \\
%	\sigma_{em} &= \ket{e} \bra{m}.
\eea
where $\mu, \nu \in \{g, e\}$.
The Heisenberg equations of motion for these operators become
\be
\frac{d}{d t} \sigma_{\mu \nu} = i \omega_{\mu \nu} \sigma_{\mu \nu} - \xi_{\mu \nu} - i \zeta_{\mu \nu} a_{\rm in}(t) + i a_{\rm in}^\dag(t) \zeta_{\nu \mu}^\dag
\label{eq:Heisenberg}
\ee
with
\bea
\xi_{\mu \nu} =&\ \frac{1}{2} \sigma_{\mu \nu} \sigma_t^\dag \sigma_t + \frac{1}{2} \sigma_t^\dag \sigma_t \sigma_{\mu \nu} - \sigma_t^\dag \sigma_{\mu \nu} \sigma_t \label{eq:ximunu} \\
\zeta_{\mu\nu} =&\ \comm{\sigma_{\mu \nu}}{\sigma_t^\dag}.
\eea
The derivation of \eqref{eq:Heisenberg}, which closely follows that for a three-level atom in an open waveguide in Ref.~\cite{Koshino2013}, is given in \ref{sec:AppendixStrongly}. The main difference is that, in an open waveguide, the atom couples to two continua of modes in the waveguide, one right-moving and one left-moving, which both enter in \eqref{eq:Hafstrong}. This means that $\sigma_t = \sqrt{\Gamma_{10} / 2} \sigma_{01}$ in the open-waveguide case and leads to a $\xi_{\mu \nu}$ with the right-hand side in \eqref{eq:ximunu} multiplied by 2. In the end, these differences lead to a twice as large amplification with the mirror than without it.

%%%%%%%%%%%%%%%%%%%%%%%%%%

\subsection{Steady-state solution and linear response}

The expectation values of the operators $\expec{\sigma_{\mu \nu}}$ are divided into steady-state and linear-response components~\cite{Koshino2013}
\bea
\expec{\sigma_{\mu \nu}} = \expec{\sigma_{\mu \nu}}_S + \expec{\sigma_{\mu \nu}}_L e^{i(\omega_d - \omega_p)t},
\label{eq:linear-response}
\eea
where $\omega_p$ is the probe frequency. The steady-state component $\expec{\sigma_{\mu \nu}}_S$ of \eqref{eq:linear-response} is calculated from \eqref{eq:Heisenberg} with the probe turned off ($\expec{a_{\rm in}} = 0$), i.e., by solving
\be
i \omega_{\mu \nu} \expec{\sigma_{\mu \nu}}_S - \sum_{\mu' \nu'} \xi_{\mu\nu,\mu'\nu'} \expec{\sigma_{\mu' \nu'}}_S = 0,
\label{eq:E1}
\ee
where $\xi_{\mu \nu, \mu' \nu'} = \brakket{\mu'}{\xi_{\mu,\nu}}{\nu'}$, and applying the condition $\sum_{\mu} \expec{\sigma_{\mu \mu}}_S = 1$. Inserting them into the following equations (see \ref{sec:AppendixStrongly}), we can calculate the linear response components by solving
\be
i \mleft( \omega_{\mu \nu} + \omega_p - \omega_d \mright) \expec{\sigma_{\mu \nu}}_L - \sum_{\mu' \nu'} \xi_{\mu \nu, \mu' \nu'} \expec{\sigma_{\mu' \nu'}}_L = i F \times \sum_{\mu' \nu'} \zeta_{\mu \nu, \mu' \nu'} \expec{\sigma_{\mu' \nu'}}_S
\label{eq:E2}
\ee
with $\zeta_{\mu \nu, \mu' \nu'} = \brakket{\mu'}{\zeta_{\mu \nu}}{\nu'}$ and $F$ the amplitude of the weak probe ($\abssq{F}$ is the number of incoming probe photons per second).

%%%%%%%%%%%%%%%%%%%%%%%%%%

\subsection{Amplification}

The reflection coefficient is defined by
\be
r = \frac{\expec{a_{\text{out}}}}{\expec{a_\text{in}}},
\ee
with
\bea
\expec{a_{\text{in}}} =&\ F e^{i(\omega_d - \omega_p)t} \\
\expec{a_{\text{out}}} =&\ \expec{a_{\rm in}} - i \expec{\sigma_t} = \mleft( F - i \sum_{\mu \nu} \sigma_{t, \mu \nu} \expec{\sigma_{\mu \nu}}_L \mright) e^{i(\omega_d - \omega_p)t}
\eea
and $\sigma_{t, \mu \nu} = \brakket{\mu}{\sigma_t}{\nu}$. For the two-level system with $\delta \omega_{10} = 0$, the reflection coefficient becomes
\be
r =	1-\frac{2 \Gamma_{10}^{2}\left(\Gamma_{10}^{3}-3 i\Gamma_{10}^{2} \delta-2 \Gamma_{10} \delta^{2}+2 i \delta \Omega_d^{2}\right)}{(\Gamma_{10}-2 i\delta)\left(\Gamma_{10}^{2}+2 \Omega_d^{2}\right)\left(\Gamma_{10}^{2}-3 i \Gamma_{10} \delta-2 \delta^{2}+2 \Omega_d^{2}\right)},
%r =	\frac{-\Gamma_{10}^2 \mleft( \Gamma_{10} - i \delta \mright)\mleft( \Gamma_{10} - 2 i \delta \mright) \mleft( \Gamma_{10} + 2 i \delta \mright) + 2 \mleft( 2 \Gamma_{10}^3 - 9 i \Gamma_{10}^2 \delta - 8 \Gamma_{10} \delta^2 + 4 i \delta^3 \mright) \Omega_d^2 + 4 \mleft( \Gamma_{10} - 2 i \delta \mright) \Omega_d^4}{\mleft( \Gamma_{10} - 2 i \delta \mright) \mleft( \Gamma_{10}^2 + 2 \Omega_d^2 \mright) \mleft( \Gamma_{10}^2 - 3 i \Gamma_{10} \delta - 2 \delta^2 + 2 \Omega_d^2 \mright)},
\label{eq:r-2LS}
\ee
where $\delta = \omega_{10} - \omega_p$.

By maximizing $\abs{r}$ using \eqref{eq:r-2LS}, we find that the maximum possible amplitude gain is $\abs{r} \approx 1.069$; it is achieved for the drive amplitude $\Omega_{10} \approx 2 \Gamma_{10}$ and probe detuning of $\delta \approx \pm 1.2 \Gamma_{10}$. It is interesting to note that the experiment in Ref.~\cite{Wen2018} appears to have come very close to this theoretical maximum.

Performing a similar analysis of the reflection coefficient for a two-level system in an open waveguide, we find that the maximum reflection for the same drive amplitude and detuning is only given by $\abssq{r} \approx 1.034$, which is only half of the gain for the atom in front of a mirror. This makes sense, since the atomic output is divided between two propagation directions in the open waveguide, while it is collected in a single output channel when a mirror is included.

%===================================================
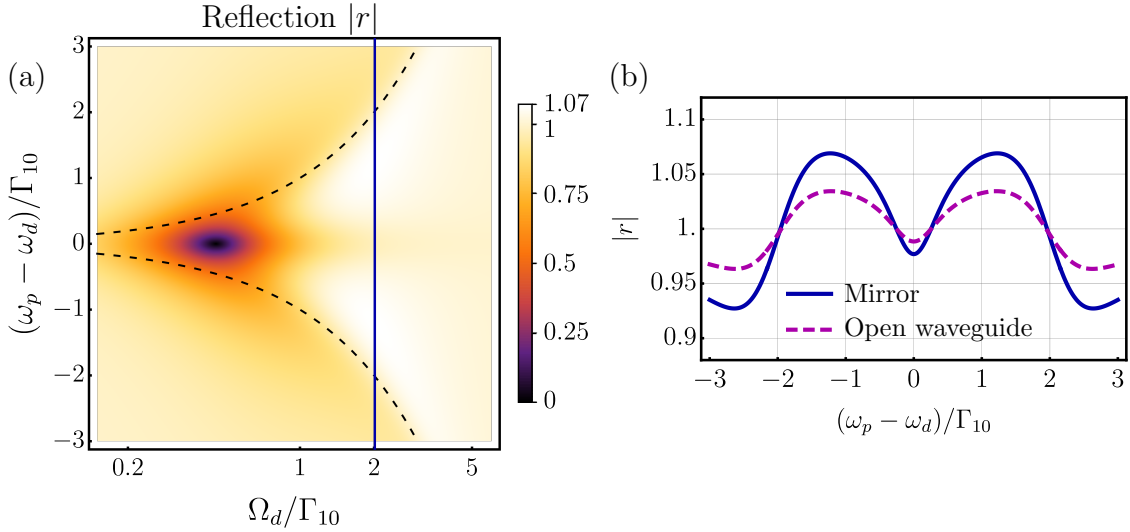
\begin{figure}[t]
	\centering
	\begin{minipage}{0.5\linewidth}
		\begin{overpic}[width=1\textwidth]{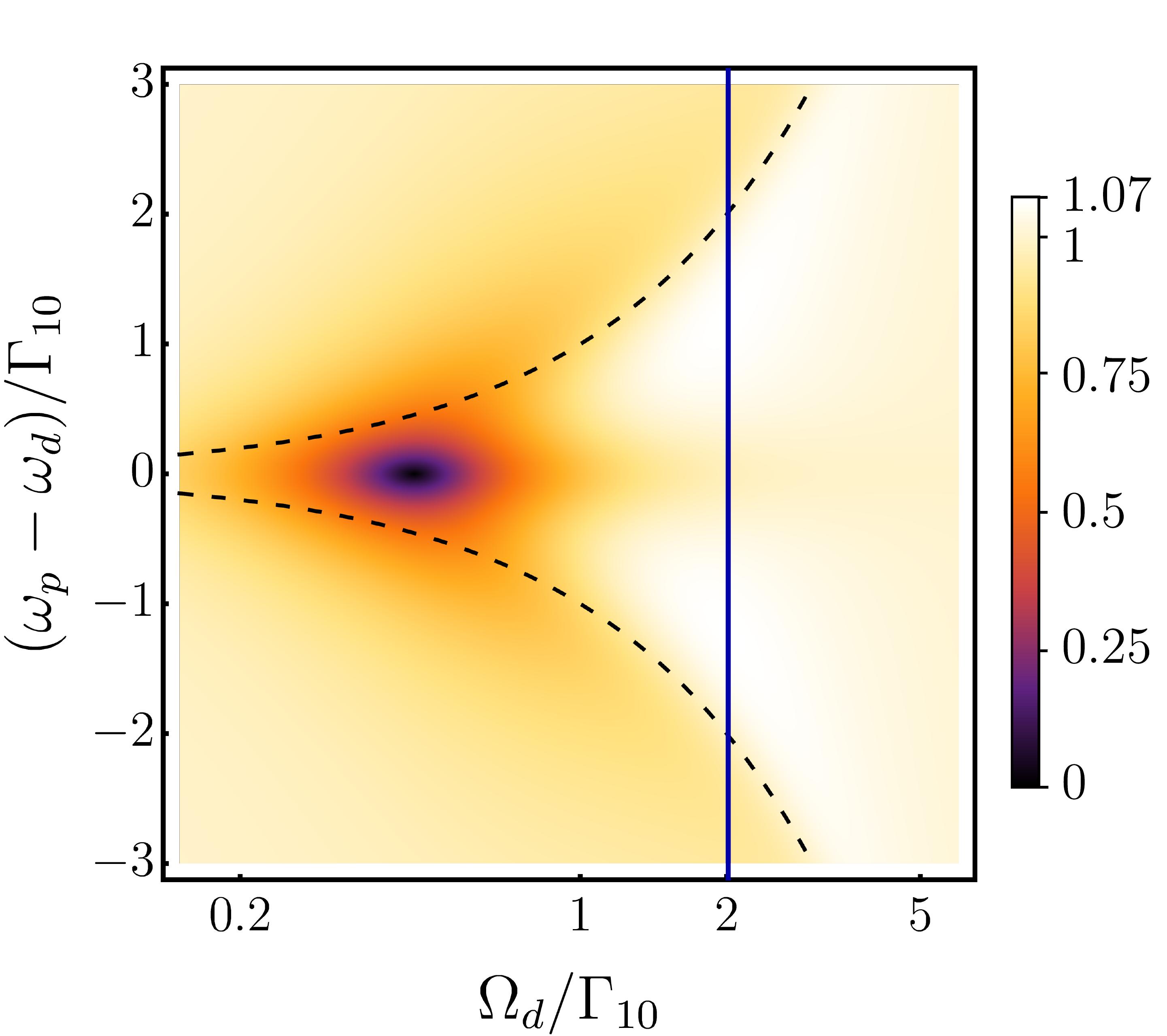}%September2020/TwoLevelMirror.nb
		    \put(33,85){Reflection $|r|$}
			\put(0,75){$\text{(a)}$}
			\put(102,75){$\text{(b)}$}
		\end{overpic}
	\end{minipage}
	\begin{minipage}{0.49\linewidth}
		\begin{overpic}[width=0.9\linewidth]{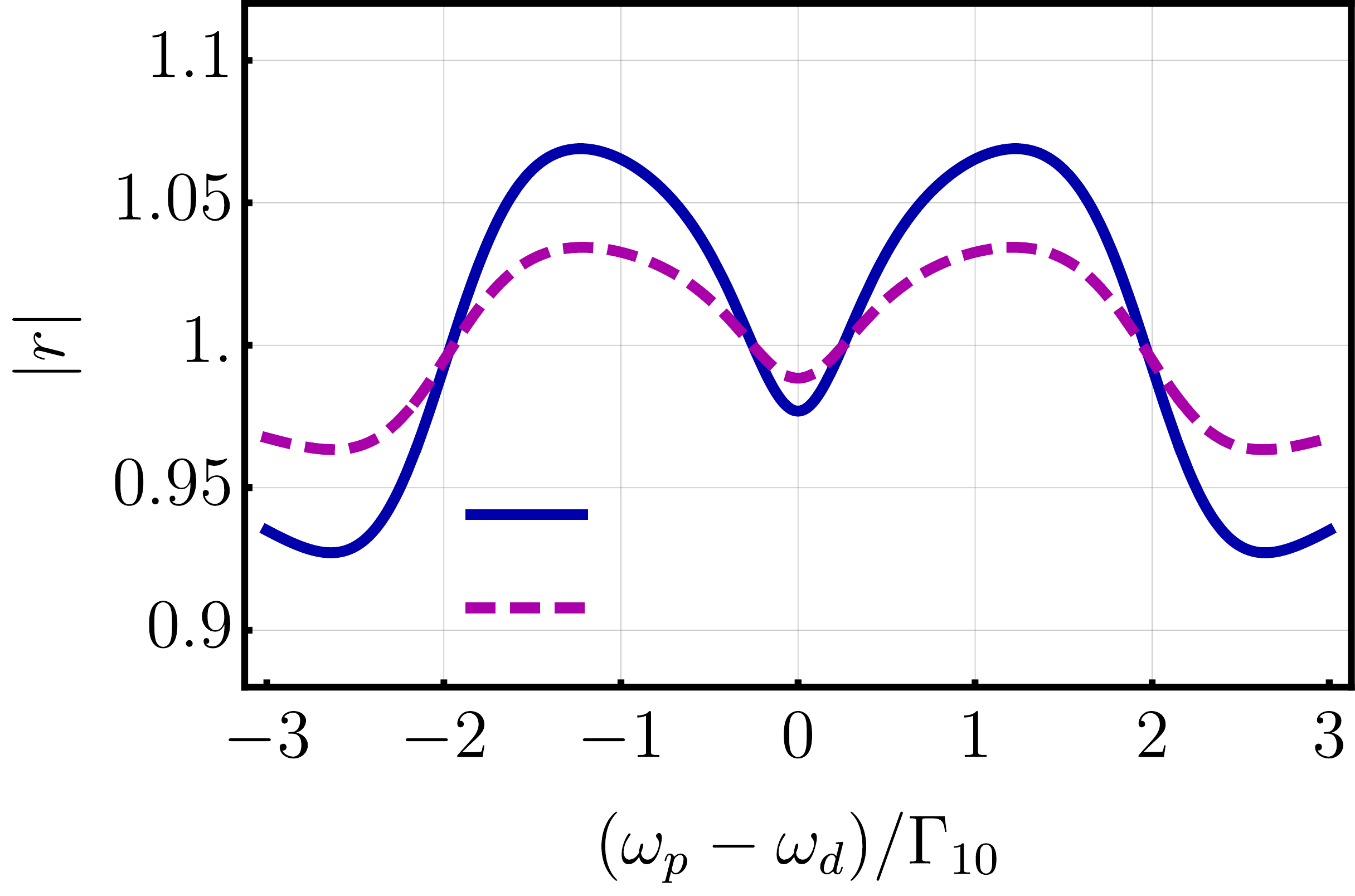} %September2020/TwoLevelMirror.nb
		\put(45.5,26.5){\footnotesize Mirror}
		\put(45.5,20){\footnotesize Open waveguide}
		\end{overpic}
	\end{minipage}
	\caption{Maximizing the reflection coefficient of a strongly driven two-level system in front of a mirror.
	(a) Reflection coefficient $\abs{r}$ of a weak probe for resonant drive ($\omega_d - \omega_{10} = 0$) as a function of the detuning of the probe frequency and drive amplitude. The maximum possible amplitude gain can be seen in the bright areas around $\Omega_d \approx 2 \Gamma_{10}$ (marked by the solid line) and $\mleft( \omega_p - \omega_{10} \mright) \approx \pm 1.2 \Gamma_{10}$.
	(b) A plot of the linecut at $\Omega_d = 2 \Gamma_{10}$ in (a). It compares the reflection of a two-level system in front of a mirror (blue, the linecut) to that in an open waveguide (purple, dashed). In both cases, we observe a maximum gain around $\mleft( \omega_p - \omega_{10} \mright) \approx \pm 1.2 \Gamma_{10}$, but the gain for the atom in front of a mirror is around \unit[6.9]{\%}, around twice the gain for the atom in an open waveguide, which is around \unit[3.4]{\%}.
	}
	\label{fig:TwoLevelStrong}
\end{figure}
%===================================================

In \figpanel{fig:TwoLevelStrong}{a}, we plot the reflection coefficient of the strongly driven two-level system as a function of the detuning $\delta$ of the probe frequency and the drive strength $\Omega_d$ for resonant drive. The bright areas correspond to gain and the dark areas to attenuation. We can see how gain is achieved at probe frequencies in-between the frequencies corresponding to the Mollow triplet. In \figpanel{fig:TwoLevelStrong}{b}, the linecut of \figpanel{fig:TwoLevelStrong}{a} is depicted. The blue curve, showing the reflection of the two-level system in front of a mirror, has a higher maximum gain than the reflection of the two-level system in an open waveguide, seen as a comparison by the purple dashed curve. This coincides with the analysis of the reflection coefficient above.

%\textcolor{red}{Expand for $\Omega > \gamma$ and compare with Mollow.}

%%%%%%%%%%%%%%%%%%%%%%%%%%%%%%%%%%%%%%%%%%%%%%%

\section{Amplification with a strongly two-photon-driven three-level atom in front of a mirror}
\label{sec:3Levels2Photon}

For our last setup, we consider a strongly driven three-level atom in front of a mirror, with the drive at half the $\ket{0} \leftrightarrow \ket{2}$ transition frequency, as sketched in \figpanel{fig:System}{d}. As shown experimentally for an open waveguide in Ref.~\cite{Koshino2013}, amplification can be achieved in this setup through population inversion among the dressed states of the three-level atom.

%%%%%%%%%%%%%%%%%%%%%%%%%%

\subsection{Hamiltonian and equations of motion}

We consider the same Hamiltonian as for the two-level case, Eqs.~(\ref{eq:Hstrong})--(\ref{eq:Hafstrong}), but including the third atomic level in the bare atomic Hamiltonian and in the Hamiltonian describing the interaction between the atom and the waveguide. The atom Hamiltonian in \eqref{eq:Ha2strong} is modified to read
\be
H_a = \delta_{10} \sigma_{11} + \delta_{20} \sigma_{20} + E \mleft( \sigma_t + \sigma_t^\dag \mright)
\ee
where $\delta_{10} = \omega_{10} - \omega_d$, $\delta_{20} = \omega_{20} - 2 \omega_d$, and $\sigma_t = \sqrt{\Gamma_{10}} \sigma_{01} + \sqrt{\Gamma_{21}} \sigma_{12}$. This new expression for $\sigma_t$ is the only change required in the interaction Hamiltonian in \eqref{eq:Hafstrong}.

We set up and solve the equations of motion for the dressed-state operators $\sigma_{\mu \nu}$ in the same way as for the two-level system in \secref{subsec:TwoLevelStrong_HEOM}, but with $\mu, \nu \in \{g, m, e \}$, where $\{g, m, e \}$ are the dressed states of the three-level system. As shown in detail in \ref{sec:AppendixStrongly}, the equations for the steady-state and linear-response components of the reflected probe signal are the same as in Eqs.~(\ref{eq:E1})--(\ref{eq:E2}) in \secref{subsec:TwoLevelStrong_HEOM}, except for the new definitions of variables given here.

%%%%%%%%%%%%%%%%%%%%%%%%%%

\subsection{Amplification}

%===================================================
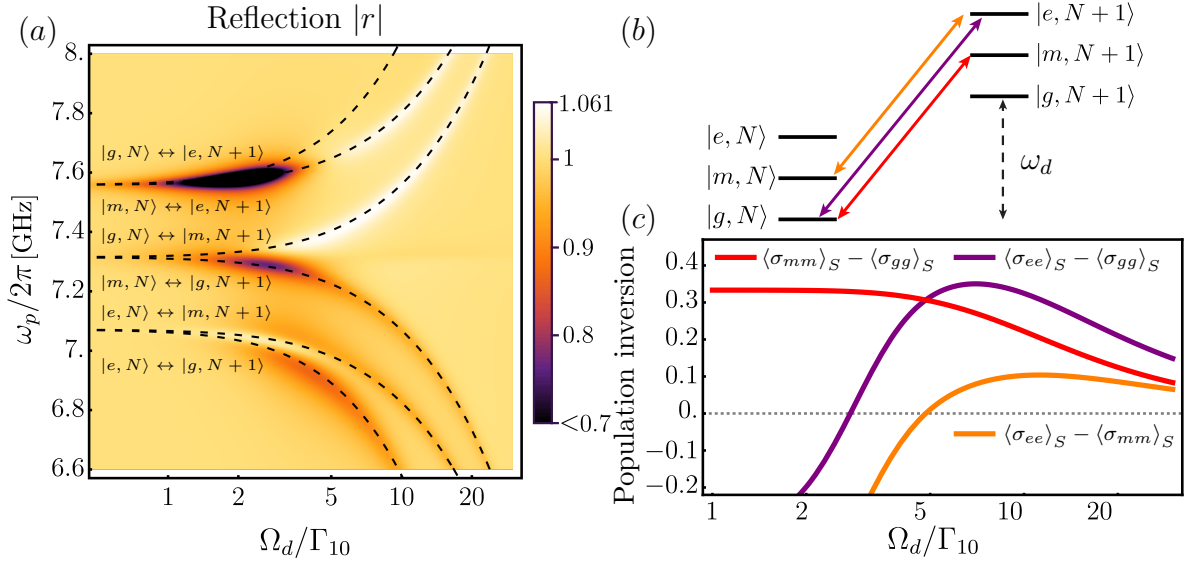
\begin{figure}[t]
	\centering
		\begin{overpic}[width=1\textwidth]{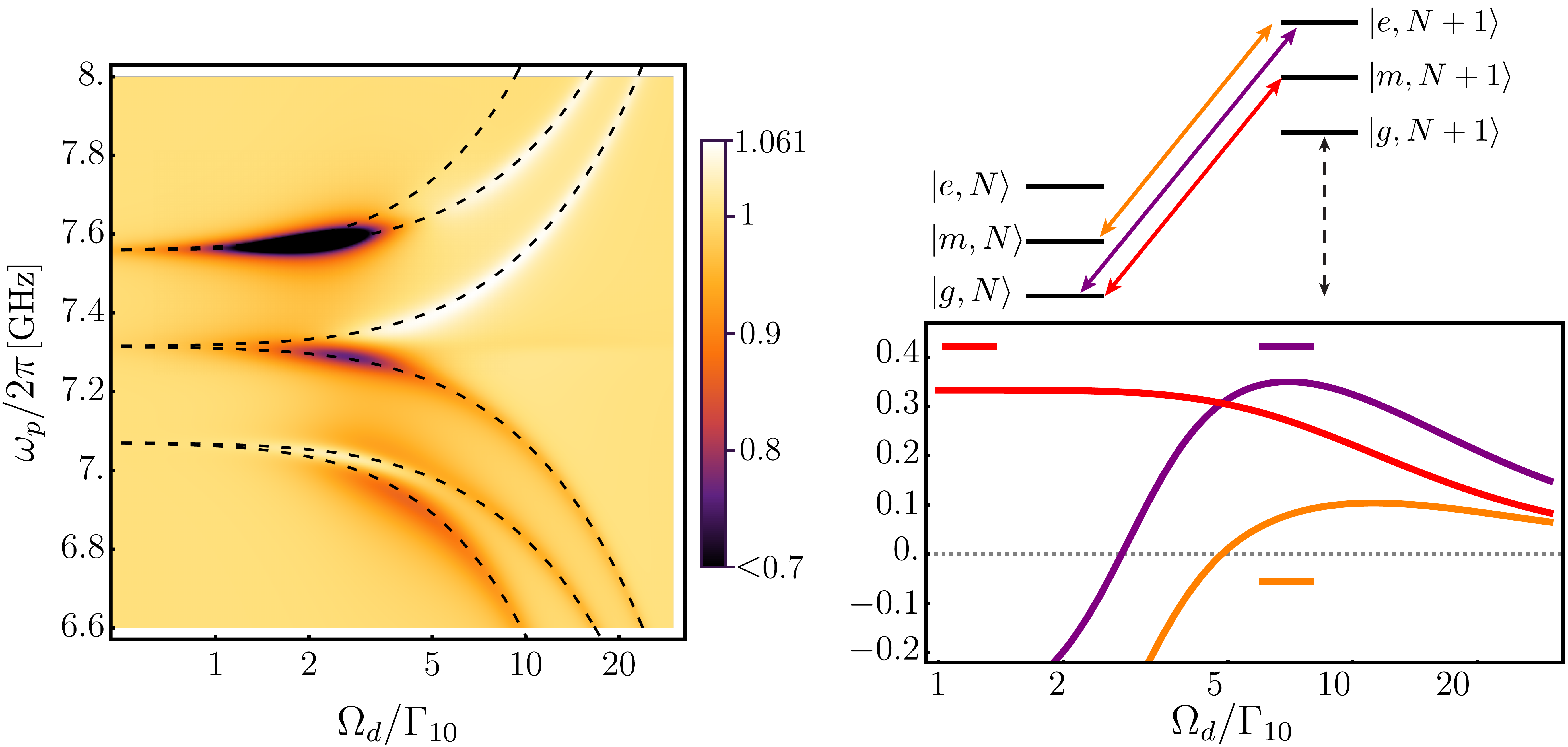}%September2020/NewPlot.nb
		    \put(1,44){$(a)$}
		    \put(52,44){$(b)$}
		    \put(52,28){$(c)$}
		    \put(17,45){Reflection $|r|$}
			\put(8,34){\tiny $\ket{g,N} \leftrightarrow \ket{e, N+1}$}
			\put(8,29.5){\tiny $\ket{m,N} \leftrightarrow \ket{e, N+1}$}
			\put(8,27){\tiny $\ket{g,N} \leftrightarrow \ket{m, N+1}$}
			\put(8,23){\tiny $\ket{m,N} \leftrightarrow \ket{g, N+1}$}
			\put(8,20.5){\tiny $\ket{e,N} \leftrightarrow \ket{m, N+1}$}
			\put(8,16){\tiny $\ket{e,N} \leftrightarrow \ket{g, N+1}$}
			\put(86,33){$\omega_d$}
			%
			%\put(59,30){\scriptsize $\ket{g,N}$}
			%\put(59,34){\scriptsize $\ket{m,N}$}
			%\put(59,38){\scriptsize $\ket{e,N}$}
			\put(51.5,4.5){\rotatebox{90}{\small Population inversion}}
			\put(64,25){\scriptsize $\expec{\sigma_{mm}}_S - \expec{\sigma_{gg}}_S$}
			\put(84.5,25){\scriptsize $\expec{\sigma_{ee}}_S - \expec{\sigma_{gg}}_S$}
			\put(84.5,10){\scriptsize $\expec{\sigma_{ee}}_S - \expec{\sigma_{mm}}_S$}
		\end{overpic}
	\caption{Amplification through population inversion among the dressed states of a strongly driven three-level atom.
	(a) Reflection as a function of the probe frequency $\omega_p$ and drive amplitude $\Omega_d$ for a drive frequency of $\omega_d / 2 \pi = \unit[7.26]{GHz}$ for a three-level system in front of a mirror with the transition frequencies $\omega_{10} / 2 \pi = \unit[7.4]{GHz}$, $\omega_{20} = 2 \omega_d$, and the relaxation rates $ \Gamma_{10} / 2 \pi = \unit[40]{MHz}$, $\Gamma_{21} = 2 \Gamma_{10}$. The dashed lines show the possible transitions between dressed states in the system. The dark region at $\omega_p / 2 \pi \approx \unit[7.58]{GHz}$ is outside of the plot range, since the reflection is low.
	(b) Sketch of the dressed states. The arrows demonstrating the transitions correspond to the three upper branches in panel (a).
	(c) Matrix elements of the steady-state solution showing the (non-)inverted population. Population inversion occurs for positive values of $\expec{\sigma_{\mu \mu}}_S - \expec{\sigma_{\nu \nu}}_S$, $\mu > \nu$, which is indicated by the grey dashed line at 0. The colors of the curves correspond to the colors of the arrows in (b).
	\label{fig:ThreeLevelStrongDensity}}
\end{figure}
%===================================================

In \figref{fig:ThreeLevelStrongDensity}, we plot the numerically computed reflection coefficient for a weak probe as a function of probe frequency $\omega_p$ and drive amplitude $\Omega_d = 2 E \sqrt{\Gamma_{10}}$ for typical experimental parameters~\cite{Koshino2013}. We observe a maximum amplitude gain of $\sim \unit[6]{\%}$. The largest gains are observed when the probe is close to resonant with one of the dressed-state transitions $\ket{m,N} \leftrightarrow \ket{e, N+1}$ and $\ket{g,N} \leftrightarrow \ket{m, N+1}$.

\begin{figure}[t]
	\centering
	\begin{overpic}[width=1\textwidth]{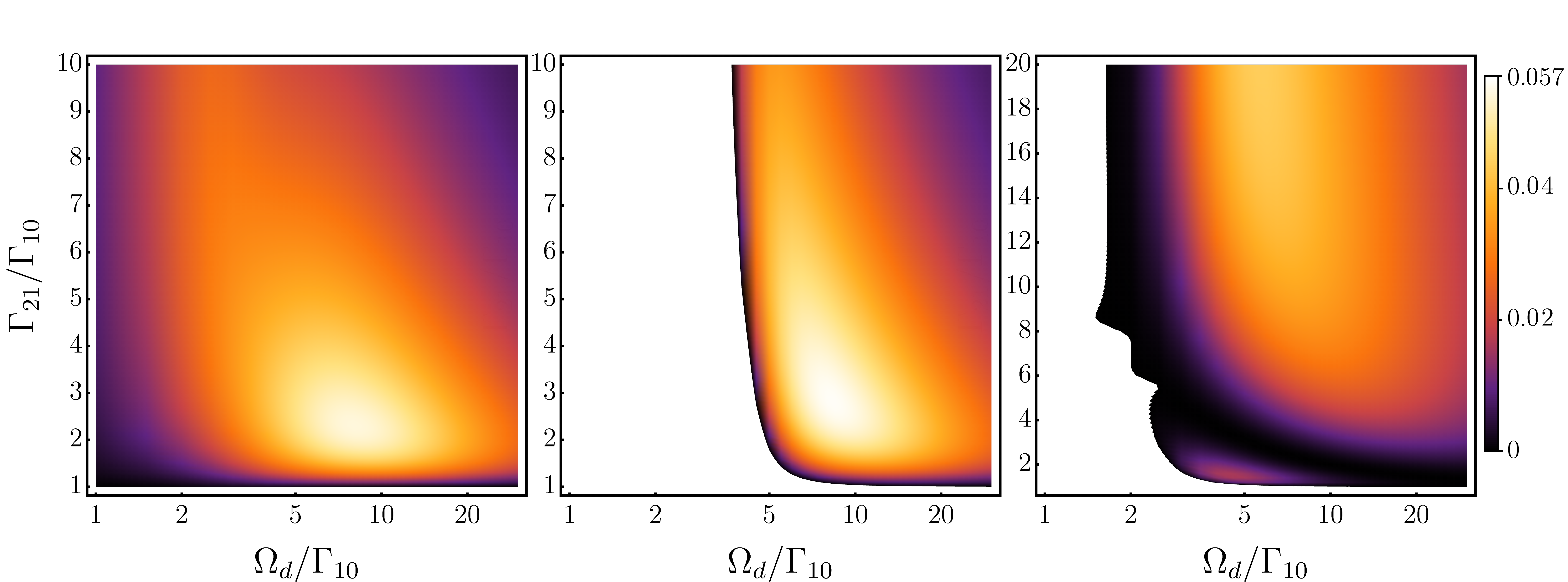}%September2020/Population.nb
		\put(9,35){\small $\ket{g,N} \leftrightarrow \ket{m, N+1}$}
		\put(39,35){\small$\ket{m,N} \leftrightarrow \ket{e, N+1}$}
		\put(71,35){\small$\ket{g,N} \leftrightarrow \ket{e, N+1}$}
		\put(3,35){(a)}
		\put(33,35){(b)}
		\put(65,35){(c)}
	\end{overpic}
	\caption{Gain of a resonant probe for the upper branches of \figref{fig:ThreeLevelStrongDensity} as a function of the drive amplitude $\Omega_d/\Gamma_{10}$ and the ratio $\Gamma_{21}/\Gamma_{10}$ between the decay rates.
	(a) Resonant gain on the $\ket{g,N}\leftrightarrow\ket{m,N+1}$ transition. We find a maximum gain of around $\unit[5.5]{\%}$ for $\Omega_d/\Gamma_{10} = 8$ and $\gamma_{21}/\gamma_{10} = 2.3$.
	(b) Resonant gain on the $\ket{m,N} \leftrightarrow \ket{e, N+1}$ transition. We find a maximum gain of around $\unit[5.7]{\%}$ for $\Omega_d/\Gamma_{10} = 8.5$ and $\Gamma_{21}/\Gamma_{10} = 2.8$. The white area in the left part of the plot corresponds to non-inverted population, i.e., ``negative gain'' (attenuation). This area corresponds to the dark spot at $\omega_p / 2 \pi \approx \unit[7.6]{GHz}$ in \figref{fig:ThreeLevelStrongDensity}.
	(c) Resonant gain on the $\ket{g,N} \leftrightarrow \ket{e, N+1}$ transition. Here, the gain increases around $\Omega_d/\Gamma_{10} = 5$ for increasing $\Gamma_{21}/\Gamma_{10}$. However, the maximum gain even for $\Gamma_{21}/\Gamma_{10} = 20$ is still smaller than the maximum gains on the other transitions.
		\label{fig:Inversion}}
\end{figure}
%===================================================

Since the population inversion among dressed states is essential for the amplification in this system, we explore further whether we can increase this population inversion by tuning the ratio between relaxation rates for the different atomic transitions. We consider the reflection along the branches for resonant probing, which are shown by the dashed black lines in \figref{fig:ThreeLevelStrongDensity}. The expression for the reflection on resonance $\omega_p = \omega_d + \omega_\nu - \omega_\mu$ can be simplified to~\cite{Koshino2013}
\be
    r = 1 + \frac{\mleft| \bra{\mu} \sigma_t \ket{\nu} \mright|^2}{\xi_{\mu \nu \mu'\nu'}} \mleft( \expec{\sigma_{\nu \nu}}_S - \expec{\sigma_{\mu \mu}}_S \mright).
    \label{eq:rResonance}
\ee
This equation shows that population inversion among the dressed states leads to a gain in the reflection, whereas we obtain attenuation for non-inverted population.

In \figref{fig:Inversion}, we plot the second term of the right-hand side in \eqref{eq:rResonance}, which corresponds to the gain, for the upper branches in \figref{fig:ThreeLevelStrongDensity}, as a function of the drive strength $\Omega_{d}$ and the ratio of the decay rates $\Gamma_{21}/\Gamma_{10}$. The bright parts of the panels in \figref{fig:Inversion} correspond to the highest resonant gains. Selecting the values for the drive strength $\Omega_{d}/\Gamma_{10}$ and the ratio of the decay rates $\Gamma_{21}/\Gamma_{10}$ that give the highest resonant gain, we find a maximum gain of $\unit[6.2]{\%}$ by searching around the resonance frequency of the $\ket{g,N} \leftrightarrow \ket{m, N+1}$ [\figpanel{fig:Inversion}{a}] transition for $\Omega_{d}/\Gamma_{10} = 8$ and $\Gamma_{21}/\Gamma_{10} = 2.3$. The corresponding gain for a transmon qubit in an open transmission line is around $\unit[3]{\%}$. Note that for a transmon in an open transmission line, $\Gamma_{21}/\Gamma_{10}$ is always $2$ (assuming a flat spectral density for the transmission line). The mirror allows us to tune $\Gamma_{21}/\Gamma_{10}$ to achieve a higher gain, but the increase is small, since the optimal ratio of relaxation rates is close to 2. Thus, the main contribution of the mirror to the increased gain is to direct all atomic output in one direction.

Repeating the same optimization for the $\ket{m,N} \leftrightarrow \ket{e, N+1}$ transition in \figpanel{fig:Inversion}{b}, we find a maximum gain of $\unit[6.1]{\%}$ for $\Omega_{d}/\Gamma_{10} = 8.5$ and $\Gamma_{21}/\Gamma_{10} = 2.8$. Once again, the optimal ratio of relaxation rates is close to 2, meaning that the improvement in gain compared to the open-transmission-line case is just a little more than a factor 2. The $\ket{g,N} \leftrightarrow \ket{e, N+1}$ transition [\figpanel{fig:Inversion}{c}] differs from the previous two in that the highest amplification is found when $\Gamma_{21} \gg \Gamma_{10}$. However, the maximum gain around this transition is smaller than that close to the other two transitions.

\section{Discussion and conclusion}

%===================================================
\begin{table}
\caption{A summary of the results in the article. We compare the highest amplitude gain we found for each of the three setups with a mirror in \figref{fig:System} to the highest amplitude gain found or observed for the same systems in an open waveguide.
\label{tab:summary}}
\renewcommand{\arraystretch}{1.2}
\renewcommand{\tabcolsep}{0.1cm}
\centering
\begin{tabular}{c | c | c | c}
Setup & 3 levels, $\omega_d = \omega_{20}$ & 2 levels, $\omega_d = \omega_{10}$ & 3 levels, $\omega_d = \omega_{20}/2$ \\ 
\hline
Schematic & \figpanelNoPrefix{fig:System}{b} & \figpanelNoPrefix{fig:System}{c} & \figpanelNoPrefix{fig:System}{d} \\
Gain with mirror & \unit[25]{\%} & \unit[6.9]{\%} & \unit[6.2]{\%} \\
Gain in open waveguide & \unit[12.5]{\%} & \unit[3.4]{\%} & \unit[3]{\%}  \\
\end{tabular}
\end{table}
%===================================================

In this article, we have investigated three different types of single-atom amplifiers, using population inversion, higher-order multi-photon processes, and hidden inversion in the dressed-state basis. For all these schemes, we compared the maximum achievable amplitude gain with the atom placed in front of a mirror versus when the atom was coupled to an open waveguide. The results are summarized in \tabref{tab:summary}. We note that first setup with a mirror reached an amplitude gain not too far from the absolute theoretical limit of $\sqrt{2}$, which corresponds to perfect population inversion and perfect stimulated emission.

We found that for all schemes, the gain is enhanced by the mirror, mainly because of two reasons:
\begin{enumerate}
    \item The mirror reduces the number of output channels for the electromagnetic field from two in an open waveguide to one. All output from the atom is thus contributing to the gain instead of only half.
    \item The mirror creates standing waves of the electromagnetic field through interference such that the strength of the coupling between the atom and the field becomes sensitive to the atomic position and transition frequencies. This makes it possible to tune the ratio of decay rates for different atomic transitions to increase population inversion and thus enhance amplification.
\end{enumerate}

These insights are ready to be demonstrated in experiments with superconducting qubits in waveguide QED. Specifically, we showed that our set-ups can be implemented with a transmon coupled to a 1D semi-infinite transmission line with currently available technology (at least two such experiments~\cite{Koshino2013, Wen2018} have already come close to the limits in \tabref{tab:summary}). We believe that this can prove useful for on-chip amplification to improve signal-to-noise ratios in experiments in quantum information and quantum optics.

An important direction for future work is to investigate how the achievable gain changes if more qubits are added to the setups described here. For example, one could imagine a cascaded setup of atoms in front of mirrors with circulators ensuring unidirectional propagation from one mirror to the next, as shown to enhance photon detection with three-level atoms in Ref.~\cite{Sathyamoorthy2014}. It could also be interesting to check how the anharmonicity of the qubit affects the gain for the three-level system with a two-photon drive. 
%Future work: does anharmonicity impact the performance of the last setup? We didn't really check that... Is there an argument for why it should or shouldn't have an impact?

%%%%%%%%%%%%%%%%%%%%%%%%%%%%%%%%%%%%%%%%%%%%%%%

\section*{Acknowledgements}

%We acknowledge useful discussions with XXX, YYY, and ZZZ.
%We also thank AAA, BBB, and CCC for comments on the manuscript.
EW acknowledges funding from the Swedish Research Council (VR) through Grant No. 2016-06059.
ICH acknowledges financial support from the MOST of Taiwan under project 109-2636-M-007-007 and the Center for Quantum Technology from the Featured Areas Research Center Program within the framework of the Higher Education Sprout Project by the Ministry of Education (MOE) in Taiwan.
% We should acknowledge the grant that Io-Chun and I have, that paid our trip to Taiwan
AFK and PD acknowledge support from the Knut and Alice Wallenberg Foundation through the Wallenberg Centre for Quantum Technology (WACQT). AFK acknowledges support from the Swedish Research Council (grant number 2019-03696).

%------------------------------------------ APPENDIX ----------------------------------------

\appendix

\section{Equations of motion for a strongly driven atom in front of a mirror}
\label{sec:AppendixStrongly}

In this appendix, we derive the equations of motion for the strongly driven atom in front of a mirror. The derivation applies both for the two-level atom in \secref{sec:TwoLevelStrong} and three-level atom in \secref{sec:3Levels2Photon}. We start with the diagonalised form of the two-level $H_2^{\text{dressed}}$ and three-level Hamiltonian $H_3^{\text{dressed}}$, given by
\bea
H_a^2 =&\ \omega_g \sigma_{gg} + \omega_e \sigma_{ee}, \\ 
H_a^3 =&\ \omega_g \sigma_{gg} + \omega_m \sigma_{mm} + \omega_e \sigma_{ee}.
\eea
The field and interaction Hamiltonians are
\bea
H_f &=\ \int d \omega \ \omega a^\dag (\omega) a (\omega), \\
H_{a-f} &=\ \frac{1}{\sqrt{2 \pi}} \int_0^\infty d \omega \mleft( a^\dag (\omega) \sigma_t + \sigma_t^\dag a (\omega) \mright),
\eea
with $\sigma_t = \sqrt{\Gamma_{10}} \sigma_{01}$ for the two-level system, and $\sigma_t = \sqrt{\Gamma_{10}} \sigma_{01} + \sqrt{\Gamma_{21}} \sigma_{12}$ for the three-level system.

In the following, we denote the dressed-state operators by $\sigma_{\mu \nu}$ and the dressed-state transition frequencies by $\omega_{\mu \nu} = \omega_\mu - \omega_\nu$, with $\mu \nu \in \{ g, m, e \}$ for the three-level atom and $\mu \nu \in \{ g, e \}$ for the two-level atom. The equation of motion for $\sigma_{\mu \nu}$ can be calculated by the Heisenberg equation
\be
\frac{d }{d t} \sigma_{\mu \nu} = i \comm{H}{\sigma_{\mu \nu}}.
\ee
We find
\bea
i \comm{H_a}{\sigma_{\mu \nu}} &= i \comm{\omega_g \sigma_{gg} + \omega_m \sigma_{mm} + \omega_e \sigma_{ee}}{\sigma_{\mu \nu}} \nn\\
&= \omega_g \mleft( \ket{g} \overbrace{\braket{g}{\mu}}^{\delta_{g\mu}}\bra{\nu} - \ket{\mu} \overbrace{\braket{\nu}{g}}^{\delta_{\nu g}} \bra{g} \mright) \nn\\
&+ \omega_m \mleft( \ket{m} \overbrace{\braket{m}{\mu}}^{\delta_{m\mu}}\bra{\nu} - \ket{\mu} \overbrace{\braket{\nu}{m}}^{\delta_{\nu m}} \bra{m} \mright) \nn\\
&+ \omega_e \mleft( \ket{e} \overbrace{\braket{e}{\mu}}^{\delta_{e\mu}}\bra{\nu} - \ket{\mu} \overbrace{\braket{\nu}{e}}^{\delta_{\nu e}} \bra{e} \mright) \nn\\
&= i \mleft( \omega_\mu - \omega_\nu \mright) \sigma_{\mu \nu}, \label{eq:EoMsigmat1}
\eea
and
\bea
i &\comm{H_{a-f}}{\sigma_{\mu \nu}} = \frac{i}{\sqrt{2 \pi}} \int_0^\infty d \omega \mleft( a^\dag (\omega) \comm{\sigma_t}{\sigma_{\mu \nu}} + \comm{\sigma_t^\dag}{\sigma_{\mu \nu}} a (\omega) \mright) \nn\\
&= \frac{i}{\sqrt{2 \pi}} \int_0^\infty d \omega a^\dag (\omega) \comm{\sigma_t}{\sigma_{\mu \nu}} + \frac{i}{\sqrt{2 \pi}} \int_0^\infty d \omega \comm{\sigma_t^\dag}{\sigma_{\mu \nu}} a (\omega).
\label{eq:af}
\eea
In the same way, we calculate the equation of motion for $a(\omega)$:
\bea
\dot{a} (\omega) &=\ i \int_{-\infty}^\infty \diff \omega \omega' \comm{a^\dag(\omega) a(\omega)}{a(\omega')} + i \int_{-\infty}^\infty \diff \omega \sqrt{\frac{1}{2 \pi}} \comm{a^\dag (\omega) \sigma_t}{a (\omega')} \nn\\
&= - i \omega a(\omega) - \frac{i}{\sqrt{2 \pi}} \sigma_t.
\label{eq:EOMa}
\eea
The solution of \eqref{eq:EOMa} can be written as
\be
a(\omega) = e^{-i \omega \mleft( t - t_0 \mright)} a_0(\omega) - \frac{i}{\sqrt{2 \pi}} \int_{t_0}^t d t' \sigma_t \mleft( t' \mright) e^{- i \omega \mleft( t - t' \mright)}.
%a^\dag(\omega) &=& e^{i \omega \mleft( t - t_0 \mright)} a^\dag_0(\omega) + \frac{i}{\sqrt{2 \pi}} \int_{t_{0}}^t d t' \sigma^\dag_t \mleft( t' \mright) e^{i \omega \mleft( t - t' \mright)}.
\ee
Inserting this solution into \eqref{eq:af}, we find
\bea
\frac{i}{\sqrt{2 \pi}} \int_0^\infty d \omega a^\dag(\omega) \comm{\sigma_t}{\sigma_{\mu \nu}} =&\ \frac{i}{\sqrt{2 \pi}} \int_0^\infty d \omega  e^{i \omega \mleft( t - t_0 \mright)} a^\dag_0(\omega) \comm{\sigma_t}{\sigma_{\mu \nu}} \nn\\
&- \frac{1}{2\pi} \int_{t_0}^t d t' \int_0^\infty d \omega e^{i \omega \mleft( t - t_0 \mright)} \sigma_t^\dag \mleft( t' \mright) \comm{\sigma_t}{\sigma_{\mu \nu}} \nn\\
&= - i a_{\text{in}}^\dag \comm{\sigma_t}{\sigma_{\mu \nu}} - \int_{t_0}^t d t' \delta \mleft( t - t' \mright) \sigma_t^\dag \mleft( t' \mright) \comm{\sigma_t}{\sigma_{\mu \nu}} \nn\\
&= - i a_{\text{in}}^\dag \comm{\sigma_t}{\sigma_{\mu \nu}} - \frac{1}{2} \sigma_t^\dag \comm{\sigma_t}{\sigma_{\mu \nu}},
\eea
and
\bea
\frac{i}{\sqrt{2 \pi}} \int_0^\infty d \omega a(\omega) \comm{\sigma^\dag_t}{\sigma_{\mu \nu}} =&\ \frac{i}{\sqrt{2 \pi}} \int_0^\infty d \omega e^{- i \omega \mleft( t - t_0 \mright)} a_0(\omega) \comm{\sigma^\dag_t}{\sigma_{\mu \nu}} \nn\\
&+ \frac{1}{2\pi} \int_{t_0}^t d t' \int_0^\infty d \omega e^{- i \omega \mleft( t - t_0 \mright)} \comm{\sigma_t^\dag}{\sigma_{\mu \nu}} \sigma_t \mleft( t' \mright) \nn\\
&= i \comm{\sigma_t^\dag}{\sigma_{\mu \nu}} a_{\text{in}} + \int_{t_0}^t d t' \delta \mleft( t' - t \mright) \comm{\sigma_t^\dag}{\sigma_{\mu \nu}} \sigma_t \mleft( t' \mright) \nn \\
&= i \comm{\sigma_t^\dag}{\sigma_{\mu \nu}} a_{\text{in}} + \frac{1}{2} \comm{\sigma_t^\dag}{\sigma_{\mu \nu}} \sigma_t,
\eea
where we used
\be
\int_{t_{0}}^t d t' \sigma_t \mleft( t' \mright) \delta \mleft( t - t' \mright) = \frac{1}{2} \sigma_t (t)
\ee
and defined
\be
a_{\mathrm{in}}(t) = \frac{1}{\sqrt{2 \pi}} \int_{-\infty}^\infty d \omega e^{- i \omega \mleft( t - t_0 \mright)} a_0(\omega).
%a_{\mathrm{in}}^\dagger (t)&=\frac{1}{\sqrt{2 \pi}} \int_{-\infty}^{\infty} d \omega e^{i \omega\left(t-t_{0}\right)} a^\dagger_{0}(\omega).
\ee

Combining the above results, the equation of motion for $\sigma_{\mu \nu}$ becomes
\bea
\frac{d}{d t} \sigma_{\mu \nu} &= i (\omega_\mu - \omega_\nu) \sigma_{\mu \nu} - i a_{\text{in}}^\dag \comm{\sigma_t}{\sigma_{\mu \nu}} +  \frac{1}{2} \sigma_t^\dag \comm{\sigma_t}{\sigma_{\mu \nu}} + i \comm{\sigma_t^\dag}{\sigma_{\mu \nu}} a_{\text{in}} - \frac{1}{2} \comm{\sigma_t^\dag}{\sigma_{\mu \nu}} \sigma_t \nn\\
&= i \omega_{\mu \nu} \sigma_{\mu \nu} + i a_{\text{in}}^\dag \comm{\sigma_{\mu \nu}}{\sigma_t} - i \comm{\sigma_{\mu \nu}}{\sigma_t} a_{\text{in}} - \frac{1}{2} \sigma_{\mu \nu} \sigma_t^\dag \sigma_t - \frac{1}{2} \sigma_t^\dag \sigma_t \sigma_{\mu \nu} + \sigma_t^\dag \sigma_{\mu \nu} \sigma_t \nn\\
&= i \omega_{\mu \nu} \sigma_{\mu \nu} - \xi_{\mu \nu} - i \zeta_{\mu \nu} a_{\mathrm{in}}(t) + i a_{\mathrm{in}}^\dag(t) \zeta_{\nu \mu}^\dag
\label{eq:EOMsigmamunu}
\eea
with
\bea
\xi_{\mu\nu} &= \frac{1}{2}\sigma_{\mu \nu} \sigma_t^\dag \sigma_t + \frac{1}{2} \sigma_t^\dag \sigma_t \sigma_{\mu \nu} - \sigma_t^\dag \sigma_{\mu \nu} \sigma_t \\
\zeta_{\mu\nu} &= \comm{\sigma_{\mu \nu}}{\sigma_t^\dag}.
\eea
The steady-state solution to \eqref{eq:EOMsigmamunu} with $\expec{a_{\rm in}} = 0$ is given by
\be
\frac{d}{d t} \expec{\sigma_{\mu \nu}}_S = 0 \quad \Rightarrow \quad i \omega_{\mu \nu} \expec{\sigma_{\mu \nu}}_S - \sum_{\mu' \nu'} \xi_{\mu \nu, \mu' \nu'} \expec{\sigma_{\mu' \nu'}}_S = 0.
\ee
Together with the linear-response part $\expec{\sigma_{\mu \nu}}_L e^{i (\omega_d - \omega_p)}$, this gives
\bea
\frac{d}{d t} &\mleft( \expec{\sigma_{\mu \nu}}_S + \expec{\sigma_{\mu \nu}}_L e^{i (\omega_d - \omega_p)} \mright) = i \omega_{\mu \nu} \mleft( \expec{\sigma_{\mu \nu}}_S + \expec{\sigma_{\mu \nu}}_L e^{i (\omega_d - \omega_p)} \mright) \nn\\
&- \sum_{\mu' \nu'} \xi_{\mu \nu, \mu' \nu'} \mleft( \expec{\sigma_{\mu' \nu'}}_S + \expec{\sigma_{\mu' \nu'}}_L e^{i (\omega_d - \omega_p)} \mright) \nn\\
&- i \sum_{\mu' \nu'} \zeta_{\mu \nu, \mu' \nu'} \mleft( \expec{\sigma_{\mu' \nu'}}_S + \expec{\sigma_{\mu' \nu'}}_L e^{i (\omega_d - \omega_p)} \mright) F e^{i (\omega_d - \omega_p)} \nn\\
&+ i \sum_{\mu' \nu'} \zeta_{\nu \mu,\nu' \mu'} \mleft( \expec{\sigma_{\nu' \mu'}}_S + \expec{\sigma_{\nu' \mu'}}_L e^{i (\omega_d - \omega_p)} \mright) F e^{- i (\omega_d - \omega_p)}.
\eea
Now we use $\frac{d}{d t} \expec{\sigma_{\mu \nu}}_S = 0$, $\frac{d}{d t} \expec{\sigma_{\mu \nu}}_L = \mleft( \omega_d - \omega_p \mright) \expec{\sigma_{\mu \nu}}_L$, $i \omega_{\mu \nu} \expec{\sigma_{\mu \nu}}_S = \sum_{\mu' \nu'} \xi_{\mu \nu, \mu' \nu'} \expec{\sigma_{\mu' \nu'}}_S$, $F \times \expec{\sigma_{\mu' \nu'}}_L \ll 1$, and neglect fast rotating terms. We then find
\be
i \mleft( \omega_{\mu \nu} - \omega_d + \omega_p \mright) \expec{\sigma_{\mu \nu}}_L - \sum_{\mu' \nu'} \xi_{\mu \nu, \mu' \nu'} \expec{\sigma_{\mu \nu}}_L e^{i (\omega_d - \omega_p)} = i F \times \sum_{\mu' \nu'} \zeta_{\mu \nu, \mu' \nu'} \expec{\sigma_{\mu' \nu'}}_S,
\ee
which is used to calculate the results for amplification in Secs.~\ref{sec:TwoLevelStrong} and \ref{sec:3Levels2Photon}.

%%%%%%%%%%%%%%%%%%%%%%%%%%%%%%%%%%%%%%%%%%%%%%%

%\section*{References}
\bibliographystyle{unsrt}
\bibliography{AmplificationDraft.bbl}

\begin{thebibliography}{10}

\bibitem{Clerk2010}
A.~A. Clerk, M.~H. Devoret, S.~M. Girvin, F.~Marquardt, and R.~J. Schoelkopf.
\newblock {Introduction to quantum noise, measurement, and amplification}.
\newblock {\em Reviews of Modern Physics}, 82:1155, 2010.

\bibitem{Aumentado2020}
J.~Aumentado.
\newblock {Superconducting Parametric Amplifiers: The State of the Art in
  Josephson Parametric Amplifiers}.
\newblock {\em IEEE Microwave Magazine}, 21:45, aug 2020.

\bibitem{Haus1962}
H.~A. Haus and J.~A. Mullen.
\newblock {Quantum Noise in Linear Amplifiers}.
\newblock {\em Physical Review}, 128:2407, 1962.

\bibitem{Caves1982}
C.~M. Caves.
\newblock {Quantum limits on noise in linear amplifiers}.
\newblock {\em Physical Review D}, 26:1817, oct 1982.

\bibitem{Leuchs2013}
Gerd Leuchs and Markus Sondermann.
\newblock {Light–matter interaction in free space}.
\newblock {\em Journal of Modern Optics}, 60:36, jan 2013.

\bibitem{Gerhardt2007}
I.~Gerhardt, G.~Wrigge, P.~Bushev, G.~Zumofen, M.~Agio, R.~Pfab, and
  V.~Sandoghdar.
\newblock {Strong Extinction of a Laser Beam by a Single Molecule}.
\newblock {\em Physical Review Letters}, 98:033601, jan 2007.

\bibitem{Vamivakas2007}
A.~N. Vamivakas, M.~Atat{\"{u}}re, J.~Dreiser, S.~T. Yilmaz, A.~Badolato, A.~K.
  Swan, B.~B. Goldberg, A.~Imamoglu, and M.~S. {\"{U}}nl{\"{u}}.
\newblock {Strong Extinction of a Far-Field Laser Beam by a Single Quantum
  Dot}.
\newblock {\em Nano Letters}, 7:2892, 2007.

\bibitem{Wrigge2008}
G.~Wrigge, I.~Gerhardt, J.~Hwang, G.~Zumofen, and V.~Sandoghdar.
\newblock {Efficient coupling of photons to a single molecule and the
  observation of its resonance fluorescence}.
\newblock {\em Nature Physics}, 4:60, jan 2008.

\bibitem{Tey2008}
M.~K. Tey, Z.~Chen, S.~A. Aljunid, B.~Chng, F.~Huber, G.~Maslennikov, and
  C.~Kurtsiefer.
\newblock {Strong interaction between light and a single trapped atom without
  the need for a cavity}.
\newblock {\em Nature Physics}, 4:924, 2008.

\bibitem{Hwang2009}
J.~Hwang, M.~Pototschnig, R.~Lettow, G.~Zumofen, A.~Renn, S.~G{\"{o}}tzinger,
  and V.~Sandoghdar.
\newblock {A single-molecule optical transistor}.
\newblock {\em Nature}, 460:76, jul 2009.

\bibitem{Leong2016}
Victor Leong, Mathias~Alexander Seidler, Matthias Steiner, Alessandro
  Cer{\`{e}}, and Christian Kurtsiefer.
\newblock {Time-resolved scattering of a single photon by a single atom}.
\newblock {\em Nature Communications}, 7:13716, dec 2016.

\bibitem{Roy2017}
Dibyendu Roy, C.~M. Wilson, and Ofer Firstenberg.
\newblock Colloquium: Strongly interacting photons in one-dimensional
  continuum.
\newblock {\em Reviews of Modern Physics}, 89:021001, May 2017.

\bibitem{Gu2017}
Xiu Gu, Anton~Frisk Kockum, Adam Miranowicz, Yu-xi Liu, and Franco Nori.
\newblock Microwave photonics with superconducting quantum circuits.
\newblock {\em Physics Reports}, 718-719:1--102, 2017.

\bibitem{You2011}
J.~Q. You and Franco Nori.
\newblock Atomic physics and quantum optics using superconducting circuits.
\newblock {\em Nature}, 474(7353):589--597, 2011.

\bibitem{Kockum2019a}
A.~F. Kockum and F.~Nori.
\newblock {Quantum Bits with Josephson Junctions}.
\newblock In F.~Tafuri, editor, {\em Fundamentals and Frontiers of the
  Josephson Effect}, pages 703--741. Springer, aug 2019.

\bibitem{Blais2020}
Alexandre Blais, Arne~L. Grimsmo, S.~M. Girvin, and Andreas Wallraff.
\newblock {Circuit Quantum Electrodynamics}.
\newblock may 2020.

\bibitem{Astafiev2007}
O.~Astafiev, K.~Inomata, A.~O. Niskanen, T.~Yamamoto, Yu.~A. Pashkin,
  Y.~Nakamura, and J.~S. Tsai.
\newblock Single artificial-atom lasing.
\newblock {\em Nature}, 449(7162):588--590, 2007.

\bibitem{Ashab2009}
S~Ashhab, J~R Johansson, A~M Zagoskin, and Franco Nori.
\newblock Single-artificial-atom lasing using a voltage-biased superconducting
  charge qubit.
\newblock {\em New Journal of Physics}, 11(2):023030, 2009.

\bibitem{You2007}
J.~Q. You, Yu-xi Liu, C.~P. Sun, and Franco Nori.
\newblock Persistent single-photon production by tunable on-chip micromaser
  with a superconducting quantum circuit.
\newblock {\em Physical Review B}, 75:104516, Mar 2007.

\bibitem{Marthaler2011}
M.~Marthaler, Y.~Utsumi, D.~S. Golubev, A.~Shnirman, and Gerd Sch\"on.
\newblock {Lasing without Inversion in Circuit Quantum Electrodynamics}.
\newblock {\em Physical Review Letters}, 107:093901, Aug 2011.

\bibitem{Kjaergaard2020}
M.~Kjaergaard, M.~E. Schwartz, J.~Braum{\"{u}}ller, P.~Krantz, J.~I.-J. Wang,
  S.~Gustavsson, and W.~D. Oliver.
\newblock {Superconducting Qubits: Current State of Play}.
\newblock {\em Annual Review of Condensed Matter Physics}, 11:369, mar 2020.

\bibitem{Blais2004}
Alexandre Blais, Ren-Shou Huang, Andreas Wallraff, S.~M. Girvin, and R.~J.
  Schoelkopf.
\newblock Cavity quantum electrodynamics for superconducting electrical
  circuits: An architecture for quantum computation.
\newblock {\em Physical Review A}, 69:062320, Jun 2004.

\bibitem{Wallraff2004}
A.~Wallraff, D.~I. Schuster, A.~Blais, L.~Frunzio, R.~S. Huang, J.~Majer,
  S.~Kumar, S.~M. Girvin, and R.~J. Schoelkopf.
\newblock Strong coupling of a single photon to a superconducting qubit using
  circuit quantum electrodynamics.
\newblock {\em Nature}, 431(7005):162--167, 2004.

\bibitem{Astafiev2010_2}
O.~Astafiev, A.~M. Zagoskin, A.~A. Abdumalikov, Yu.~A. Pashkin, T.~Yamamoto,
  K.~Inomata, Y.~Nakamura, and J.~S. Tsai.
\newblock Resonance fluorescence of a single artificial atom.
\newblock {\em Science}, 327(5967):840--843, 2010.

\bibitem{Paik2011}
H.~Paik, D.~I. Schuster, L.~S. Bishop, G.~Kirchmair, G.~Catelani, A.~P. Sears,
  B.~R. Johnson, M.~J. Reagor, L.~Frunzio, L.~I. Glazman, S.~M. Girvin, M.~H.
  Devoret, and R.~J. Schoelkopf.
\newblock {Observation of High Coherence in Josephson Junction Qubits Measured
  in a Three-Dimensional Circuit QED Architecture}.
\newblock {\em Physical Review Letters}, 107:240501, dec 2011.

\bibitem{Devoret2007}
M.H. Devoret, Steven Girvin, and Robert Schoelkopf.
\newblock Circuit-qed: How strong can the coupling between a josephson junction
  atom and a transmission line resonator be?
\newblock {\em Annalen der Physik}, 16(10‐11):767--779, 2007.

\bibitem{Bourassa2009}
J.~Bourassa, J.~M. Gambetta, A.~A. Abdumalikov, O.~Astafiev, Y.~Nakamura, and
  A.~Blais.
\newblock {Ultrastrong coupling regime of cavity QED with phase-biased flux
  qubits}.
\newblock {\em Physical Review A}, 80:032109, Sep 2009.

\bibitem{Niemczyk2010}
T.~Niemczyk, F.~Deppe, H.~Huebl, E.~P. Menzel, F.~Hocke, M.~J. Schwarz, J.~J.
  Garcia-Ripoll, D.~Zueco, T.~H{\"u}mmer, E.~Solano, A.~Marx, and R.~Gross.
\newblock Circuit quantum electrodynamics in the ultrastrong-coupling regime.
\newblock {\em Nature Physics}, 6(10):772--776, 2010.

\bibitem{Forn-Diaz2017}
P.~Forn-D{\'{i}}az, J.~J. Garc{\'{i}}a-Ripoll, B.~Peropadre, J.-L. Orgiazzi,
  M.~A. Yurtalan, R.~Belyansky, C.~M. Wilson, and A.~Lupascu.
\newblock {Ultrastrong coupling of a single artificial atom to an
  electromagnetic continuum in the nonperturbative regime}.
\newblock {\em Nature Physics}, 13:39, 2017.

\bibitem{Yoshihara2017}
Fumiki Yoshihara, Tomoko Fuse, Sahel Ashhab, Kosuke Kakuyanagi, Shiro Saito,
  and Kouichi Semba.
\newblock Superconducting qubit--oscillator circuit beyond the
  ultrastrong-coupling regime.
\newblock {\em Nature Physics}, 13(1):44--47, 2017.

\bibitem{Kockum2019}
A.~F. Kockum, A.~Miranowicz, S.~{De Liberato}, S.~Savasta, and F.~Nori.
\newblock {Ultrastrong coupling between light and matter}.
\newblock {\em Nature Reviews Physics}, 1:19, jan 2019.

\bibitem{Forn-Diaz2019}
P.~Forn-D{\'{i}}az, L.~Lamata, E.~Rico, J.~Kono, and E.~Solano.
\newblock {Ultrastrong coupling regimes of light-matter interaction}.
\newblock {\em Reviews of Modern Physics}, 91:025005, jun 2019.

\bibitem{Wen2018}
P.~Y. Wen, A.~F. Kockum, H.~Ian, J.~C. Chen, F.~Nori, and I.-C. Hoi.
\newblock {Reflective Amplification without Population Inversion from a
  Strongly Driven Superconducting Qubit}.
\newblock {\em Physical Review Letters}, 120:063603, Feb 2018.

\bibitem{Wu1977}
F.~Y. Wu, S.~Ezekiel, M.~Ducloy, and B.~R. Mollow.
\newblock {Observation of Amplification in a Strongly Driven Two-Level Atomic
  System at Optical Frequencies}.
\newblock {\em Physical Review Letters}, 38:1077--1080, May 1977.

\bibitem{Xu2007}
Xiaodong Xu, Bo~Sun, Paul~R. Berman, Duncan~G. Steel, Allan~S. Bracker, Dan
  Gammon, and L.~J. Sham.
\newblock Coherent optical spectroscopy of a strongly driven quantum dot.
\newblock {\em Science}, 317(5840):929--932, 2007.

\bibitem{Astafiev2010}
O.~V. Astafiev, A.~A. Abdumalikov, A.~M. Zagoskin, Yu.~A. Pashkin, Y.~Nakamura,
  and J.~S. Tsai.
\newblock {Ultimate On-Chip Quantum Amplifier}.
\newblock {\em Physical Review Letters}, 104:183603, May 2010.

\bibitem{Hoi2011}
I.-C. Hoi, C.~M. Wilson, G.~Johansson, T.~Palomaki, B.~Peropadre, and
  P.~Delsing.
\newblock {Demonstration of a Single-Photon Router in the Microwave Regime}.
\newblock {\em Physical Review Letters}, 107:073601, 2011.

\bibitem{Hoi2012}
I.-C. Hoi, T.~Palomaki, J.~Lindkvist, G.~Johansson, P.~Delsing, and C.~M.
  Wilson.
\newblock {Generation of Nonclassical Microwave States Using an Artificial Atom
  in 1D Open Space}.
\newblock {\em Physical Review Letters}, 108:263601, 2012.

\bibitem{vanLoo2013}
Arjan~F. van Loo, Arkady Fedorov, Kevin Lalumi{\`e}re, Barry~C. Sanders,
  Alexandre Blais, and Andreas Wallraff.
\newblock Photon-mediated interactions between distant artificial atoms.
\newblock {\em Science}, 342(6165):1494--1496, 2013.

\bibitem{Koshino2013}
K.~Koshino, H.~Terai, K.~Inomata, T.~Yamamoto, W.~Qiu, Z.~Wang, and
  Y.~Nakamura.
\newblock {Observation of the Three-State Dressed States in Circuit Quantum
  Electrodynamics}.
\newblock {\em Physical Review Letters}, 110:263601, Jun 2013.

\bibitem{Hoi2013}
Io-Chun Hoi, C~M Wilson, G{\"o}ran Johansson, Joel Lindkvist, Borja Peropadre,
  Tauno Palomaki, and Per Delsing.
\newblock Microwave quantum optics with an artificial atom in one-dimensional
  open space.
\newblock {\em New Journal of Physics}, 15(2):025011, 2013.

\bibitem{Hoi2013_2}
Io-Chun Hoi, Anton~F. Kockum, Tauno Palomaki, Thomas~M. Stace, Bixuan Fan, Lars
  Tornberg, Sankar~R. Sathyamoorthy, G\"oran Johansson, Per Delsing, and C.~M.
  Wilson.
\newblock {Giant Cross--Kerr Effect for Propagating Microwaves Induced by an
  Artificial Atom}.
\newblock {\em Physical Review Letters}, 111:053601, Aug 2013.

\bibitem{Liu2017}
Y.~Liu and A.~A. Houck.
\newblock {Quantum electrodynamics near a photonic bandgap}.
\newblock {\em Nature Physics}, 13:48, 2017.

\bibitem{Mirhosseini2018}
M.~Mirhosseini, E.~Kim, V.~S. Ferreira, M.~Kalaee, A.~Sipahigil, A.~J. Keller,
  and O.~Painter.
\newblock {Superconducting metamaterials for waveguide quantum
  electrodynamics}.
\newblock {\em Nature Communications}, 9:3706, 2018.

\bibitem{Sundaresan2019}
N.~M. Sundaresan, R.~Lundgren, G.~Zhu, A.~V. Gorshkov, and A.~A. Houck.
\newblock {Interacting Qubit-Photon Bound States with Superconducting
  Circuits}.
\newblock {\em Physical Review X}, 9:011021, 2019.

\bibitem{Wen2019}
P.~Y. Wen, K.-T. Lin, A.~F. Kockum, B.~Suri, H.~Ian, J.~C. Chen, S.~Y. Mao,
  C.~C. Chiu, P.~Delsing, F.~Nori, G.-D. Lin, and I.-C. Hoi.
\newblock {Large Collective Lamb Shift of Two Distant Superconducting
  Artificial Atoms}.
\newblock {\em Physical Review Letters}, 123:233602, dec 2019.

\bibitem{Mirhosseini2019}
M.~Mirhosseini, E.~Kim, X.~Zhang, A.~Sipahigil, P.~B. Dieterle, A.~J. Keller,
  A.~Asenjo-Garcia, D.~E. Chang, and O.~Painter.
\newblock {Cavity quantum electrodynamics with atom-like mirrors}.
\newblock {\em Nature}, 569:692, may 2019.

\bibitem{Kannan2020}
B.~Kannan, M.~J. Ruckriegel, D.~L. Campbell, A.~F. Kockum, J.~Braum{\"{u}}ller,
  D.~K. Kim, M.~Kjaergaard, P.~Krantz, A.~Melville, B.~M. Niedzielski,
  A.~Veps{\"{a}}l{\"{a}}inen, R.~Winik, J.~L. Yoder, F.~Nori, T.~P. Orlando,
  S.~Gustavsson, and W.~D. Oliver.
\newblock {Waveguide quantum electrodynamics with superconducting artificial
  giant atoms}.
\newblock {\em Nature}, 583:775, jul 2020.

\bibitem{Vadiraj2020}
A.~M. Vadiraj, A.~Ask, T.~G. McConkey, I.~Nsanzineza, C.~W. {Sandbo Chang},
  A.~F. Kockum, and C.~M. Wilson.
\newblock {Engineering the Level Structure of a Giant Artificial Atom in
  Waveguide Quantum Electrodynamics}.
\newblock mar 2020.

\bibitem{Johansson2009}
J.~R. Johansson, G.~Johansson, C.~M. Wilson, and F.~Nori.
\newblock {Dynamical Casimir Effect in a Superconducting Coplanar Waveguide}.
\newblock {\em Physical Review Letters}, 103:147003, 2009.

\bibitem{Wilson2011}
C.~M. Wilson, G.~Johansson, A.~Pourkabirian, M.~Simoen, J.~R. Johansson,
  T.~Duty, F.~Nori, and P.~Delsing.
\newblock {Observation of the dynamical Casimir effect in a superconducting
  circuit}.
\newblock {\em Nature}, 479:376, 2011.

\bibitem{Hoi2015}
I.~C. Hoi, A.~F. Kockum, L.~Tornberg, A.~Pourkabirian, G.~Johansson,
  P.~Delsing, and C.~M. Wilson.
\newblock {Probing the quantum vacuum with an artificial atom in front of a
  mirror}.
\newblock {\em Nature Physics}, 11(12):1045--1049, 2015.

\bibitem{Eschner2001}
J.~Eschner, C.~Raab, F.~Schmidt-Kaler, and R.~Blatt.
\newblock {Light interference from single atoms and their mirror images}.
\newblock {\em Nature}, 413:495, 2001.

\bibitem{Wilson2003}
M.~A. Wilson, P.~Bushev, J.~Eschner, F.~Schmidt-Kaler, C.~Becher, R.~Blatt, and
  U.~Dorner.
\newblock {Vacuum-Field Level Shifts in a Single Trapped Ion Mediated by a
  Single Distant Mirror}.
\newblock {\em Physical Review Letters}, 91:213602, 2003.

\bibitem{Dubin2007}
F.~Dubin, D.~Rotter, M.~Mukherjee, C.~Russo, J.~Eschner, and R.~Blatt.
\newblock {Photon Correlation versus Interference of Single-Atom Fluorescence
  in a Half-Cavity}.
\newblock {\em Physical Review Letters}, 98:183003, 2007.

\bibitem{Lu2020}
Y.~Lu, A.~Bengtsson, J.~J. Burnett, E.~Wiegand, B.~Suri, P.~Krantz, A.~F.
  Roudsari, A.~F. Kockum, S.~Gasparinetti, G.~Johansson, and P.~Delsing.
\newblock {Characterizing decoherence rates of a superconducting qubit by
  direct microwave scattering}.
\newblock {\em npj Quantum Information}, dec 2020.

\bibitem{Scigliuzzo2020}
Marco Scigliuzzo, Andreas Bengtsson, Jean-Claude Besse, Andreas Wallraff, Per
  Delsing, and Simone Gasparinetti.
\newblock {Primary thermometry of propagating microwaves in the quantum
  regime}.
\newblock {\em Physical Review X}, mar 2020.

\bibitem{Meschede1990}
D.~Meschede, W.~Jhe, and E.~A. Hinds.
\newblock {Radiative properties of atoms near a conducting plane: An old
  problem in a new light}.
\newblock {\em Physical Review A}, 41:1587, 1990.

\bibitem{Dorner2002}
U.~Dorner and P.~Zoller.
\newblock {Laser-driven atoms in half-cavities}.
\newblock {\em Physical Review A}, 66:023816, 2002.

\bibitem{Beige2002}
A.~Beige, J.~Pachos, and H.~Walther.
\newblock {Spontaneous emission of an atom in front of a mirror}.
\newblock {\em Physical Review A}, 66:063801, 2002.

\bibitem{Dong2009}
H.~Dong, Z.~R. Gong, H.~Ian, L.~Zhou, and C.~P. Sun.
\newblock {Intrinsic cavity QED and emergent quasinormal modes for a single
  photon}.
\newblock {\em Physical Review A}, 79:063847, 2009.

\bibitem{Koshino2012}
K.~Koshino and Y.~Nakamura.
\newblock {Control of the radiative level shift and linewidth of a
  superconducting artificial atom through a variable boundary condition}.
\newblock {\em New Journal of Physics}, 14:043005, 2012.

\bibitem{Wang2012}
Y.~Wang, J.~Min{\'{a}}r, G.~H{\'{e}}tet, and V.~Scarani.
\newblock {Quantum memory with a single two-level atom in a half cavity}.
\newblock {\em Physical Review A}, 85:013823, jan 2012.

\bibitem{Tufarelli2013}
T.~Tufarelli, F.~Ciccarello, and M.~S. Kim.
\newblock {Dynamics of spontaneous emission in a single-end photonic
  waveguide}.
\newblock {\em Physical Review A}, 87:013820, 2013.

\bibitem{Fang2015}
Y.-L.~L. Fang and H.~U. Baranger.
\newblock {Waveguide QED: Power spectra and correlations of two photons
  scattered off multiple distant qubits and a mirror}.
\newblock {\em Physical Review A}, 91:053845, 2015.

\bibitem{Shi2015}
T.~Shi, D.~E. Chang, and J.~I. Cirac.
\newblock {Multiphoton-scattering theory and generalized master equations}.
\newblock {\em Physical Review A}, 92:053834, nov 2015.

\bibitem{Pichler2016}
H.~Pichler and P.~Zoller.
\newblock {Photonic Circuits with Time Delays and Quantum Feedback}.
\newblock {\em Physical Review Letters}, 116:093601, 2016.

\bibitem{Pichler2017}
Hannes Pichler, Soonwon Choi, Peter Zoller, and Mikhail~D. Lukin.
\newblock {Universal photonic quantum computation via time-delayed feedback}.
\newblock {\em Proceedings of the National Academy of Sciences}, 114:11362, oct
  2017.

\bibitem{Wiegand2020}
E.~Wiegand, B.~Rousseaux, and G.~Johansson.
\newblock {Semiclassical analysis of dark-state transient dynamics in waveguide
  circuit QED}.
\newblock {\em Physical Review A}, 101:033801, mar 2020.

\bibitem{Wiegand2020a}
Emely Wiegand, Benjamin Rousseaux, and G{\"{o}}ran Johansson.
\newblock {Transmon in a semi-infinite high-impedance transmission line --
  appearance of cavity modes and Rabi oscillations}.
\newblock dec 2020.

\bibitem{Sargent1974}
M.~Sargent, M.~O. Scully, and W.~E. Lamb.
\newblock {\em {Laser Physics}}.
\newblock Addison-Wesley, 1974.

\bibitem{Silfvast1996}
W.~T. Silfvast.
\newblock {\em {Laser Fundamentals}}.
\newblock Cambridge University Press, 1996.

\bibitem{Mompart2000}
J~Mompart and R~Corbal{\'{a}}n.
\newblock Lasing without inversion.
\newblock {\em Journal of Optics B: Quantum and Semiclassical Optics},
  2(3):R7--R24, may 2000.

\bibitem{Haroche1972}
Serge Haroche and Francis Hartmann.
\newblock {Theory of Saturated-Absorption Line Shapes}.
\newblock {\em Physical Review A}, 6:1280--1300, Oct 1972.

\bibitem{Mollow1969}
B.~R. Mollow.
\newblock {Power Spectrum of Light Scattered by Two-Level Systems}.
\newblock {\em Physical Review}, 188:1969--1975, Dec 1969.

\bibitem{Abdumalikov2011}
A.~A. Abdumalikov, O.~V. Astafiev, Yu.~A. Pashkin, Y.~Nakamura, and J.~S. Tsai.
\newblock Dynamics of coherent and incoherent emission from an artificial atom
  in a 1d space.
\newblock {\em Phys. Rev. Lett.}, 107:043604, Jul 2011.

\bibitem{Mollow1972}
B.~R. Mollow.
\newblock {Stimulated Emission and Absorption near Resonance for Driven
  Systems}.
\newblock {\em Physical Review A}, 5:2217--2222, May 1972.

\bibitem{Friedmann1987}
H.~Friedmann and A.~D. Wilson-Gordon.
\newblock Dispersion profiles of the absorptive response of a two-level system
  interacting with two intense fields.
\newblock {\em Physical Review A}, 36:1333--1341, Aug 1987.

\bibitem{Kockum2020}
A.~F. Kockum.
\newblock {Quantum Optics with Giant Atoms - the First Five Years}.
\newblock In {\em International Symposium on Mathematics, Quantum Theory, and
  Cryptography (Mathematics for Industry, vol 33)}, pages 125--146. Springer,
  2021.

\bibitem{Kockum2014}
A.~F. Kockum, P.~Delsing, and G.~Johansson.
\newblock {Designing frequency-dependent relaxation rates and Lamb shifts for a
  giant artificial atom}.
\newblock {\em Physical Review A}, 90:013837, 2014.

\bibitem{Gustafsson2014}
M.~V. Gustafsson, T.~Aref, A.~F. Kockum, M.~K. Ekstr{\"{o}}m, G.~Johansson, and
  P.~Delsing.
\newblock {Propagating phonons coupled to an artificial atom}.
\newblock {\em Science}, 346:207, 2014.

\bibitem{Guo2017}
L.~Guo, A.~L. Grimsmo, A.~F. Kockum, M.~Pletyukhov, and G.~Johansson.
\newblock {Giant acoustic atom: A single quantum system with a deterministic
  time delay}.
\newblock {\em Physical Review A}, 95:053821, may 2017.

\bibitem{Manenti2017}
R.~Manenti, A.~F. Kockum, A.~Patterson, T.~Behrle, J.~Rahamim, G.~Tancredi,
  F.~Nori, and P.~J. Leek.
\newblock {Circuit quantum acoustodynamics with surface acoustic waves}.
\newblock {\em Nature Communications}, 8:975, dec 2017.

\bibitem{Kockum2018}
A.~F. Kockum, G.~Johansson, and F.~Nori.
\newblock {Decoherence-Free Interaction between Giant Atoms in Waveguide
  Quantum Electrodynamics}.
\newblock {\em Physical Review Letters}, 120:140404, 2018.

\bibitem{Karg2019}
T.~M. Karg, B.~Gouraud, P.~Treutlein, and K.~Hammerer.
\newblock {Remote Hamiltonian interactions mediated by light}.
\newblock {\em Physical Review A}, 99:063829, jun 2019.

\bibitem{Ask2019}
A.~Ask, M.~Ekstr{\"{o}}m, P.~Delsing, and G.~Johansson.
\newblock {Cavity-free vacuum-Rabi splitting in circuit quantum
  acoustodynamics}.
\newblock {\em Physical Review A}, 99:013840, jan 2019.

\bibitem{Gonzalez-Tudela2019}
A.~Gonz{\'{a}}lez-Tudela, C.~{S{\'{a}}nchez Mu{\~{n}}oz}, and J.~I. Cirac.
\newblock {Engineering and Harnessing Giant Atoms in High-Dimensional Baths: A
  Proposal for Implementation with Cold Atoms}.
\newblock {\em Physical Review Letters}, 122:203603, may 2019.

\bibitem{Guimond2020}
P.-O. Guimond, B.~Vermersch, M.~L. Juan, A.~Sharafiev, G.~Kirchmair, and
  P.~Zoller.
\newblock {A unidirectional on-chip photonic interface for superconducting
  circuits}.
\newblock {\em npj Quantum Information}, 6:32, dec 2020.

\bibitem{Guo2020}
L.~Guo, A.~F. Kockum, F.~Marquardt, and G.~Johansson.
\newblock {Oscillating bound states for a giant atom}.
\newblock {\em Physical Review Research}, 2:043014, oct 2020.

\bibitem{Wang2020}
X.~Wang, T.~Liu, A.~F. Kockum, H.-R. Li, and F.~Nori.
\newblock {Tunable Chiral Bound States with Giant Atoms}.
\newblock aug 2020.

\bibitem{Ask2020}
A.~Ask, Y.-L.~L. Fang, and A.~F. Kockum.
\newblock {Synthesizing electromagnetically induced transparency without a
  control field in waveguide QED using small and giant atoms}.
\newblock nov 2020.

\bibitem{Koch2007}
J.~Koch, T.~M. Yu, J.~Gambetta, A.~A. Houck, D.~I. Schuster, J.~Majer,
  A.~Blais, M.~H. Devoret, S.~M. Girvin, and R.~J. Schoelkopf.
\newblock {Charge-insensitive qubit design derived from the Cooper pair box}.
\newblock {\em Physical Review A}, 76:042319, 2007.

\bibitem{Sandberg2008}
M.~Sandberg, C.~M. Wilson, F.~Persson, T.~Bauch, G.~Johansson, V.~Shumeiko,
  T.~Duty, and P.~Delsing.
\newblock {Tuning the field in a microwave resonator faster than the photon
  lifetime}.
\newblock {\em Applied Physics Letters}, 92:203501, 2008.

\bibitem{Sathyamoorthy2014}
S.~R. Sathyamoorthy, L.~Tornberg, A.~F. Kockum, B.~Q. Baragiola, J.~Combes,
  C.~M. Wilson, T.~M. Stace, and G.~Johansson.
\newblock {Quantum Nondemolition Detection of a Propagating Microwave Photon}.
\newblock {\em Physical Review Letters}, 112:093601, 2014.

\end{thebibliography}

\end{document}